\documentclass[a4paper,11pt]{article}
\pdfoutput=1 

\usepackage{jinstpub} 
                     
\usepackage{indentfirst}
\usepackage{subcaption}
\usepackage{lineno}
\usepackage{rotating}
\usepackage{colortbl}
\usepackage{arydshln}

\title{Cosmic ray muon clustering for the MicroBooNE liquid argon time projection chamber using sMask-RCNN}
\arxivnumber{2201.05705}
\collaboration{MicroBooNE Collaboration}
\author[hh]{P.~Abratenko}
\author[n]{R.~An}
\author[d]{J.~Anthony}
\author[r]{L.~Arellano}
\author[gg]{J.~Asaadi}
\author[ee]{A.~Ashkenazi}
\author[k]{S.~Balasubramanian}
\author[k]{B.~Baller}
\author[t]{C.~Barnes}
\author[w]{G.~Barr}
% ...

% ...
\affiliation[d]{University of Cambridge, Cambridge CB3 0HE, United Kingdom}
% ...
\affiliation[k]{Fermi National Accelerator Laboratory (FNAL), Batavia, IL 60510, USA}
% ...
\affiliation[n]{Illinois Institute of Technology (IIT), Chicago, IL 60616, USA}
% ...
\affiliation[r]{The University of Manchester, Manchester M13 9PL, United Kingdom}
% ...
\affiliation[t]{University of Michigan, Ann Arbor, MI, 48109, USA}
% ...
\affiliation[w]{University of Oxford, Oxford OX1 3RH, United Kingdom}
% ...
\affiliation[ee]{Tel Aviv University, Tel Aviv, Israel, 69978}
% ...
\affiliation[gg]{University of Texas, Arlington, TX, 76019, USA}
\affiliation[hh]{Tufts University, Medford, MA, 02155, USA}
% ...

    \emailAdd{microboone\_info@fnal.gov}
\date{}

\abstract{In this article, we describe a modified implementation of Mask Region-based Convolutional Neural Networks (Mask-RCNN) for cosmic ray muon clustering in a liquid argon TPC and applied to MicroBooNE neutrino data. Our implementation of this network, called sMask-RCNN, uses sparse submanifold convolutions to increase processing speed on sparse datasets, and is compared to the original dense version in several metrics. The networks are trained to use wire readout images from the MicroBooNE liquid argon time projection chamber as input and produce individually labeled particle interactions within the image. These outputs are identified as either cosmic ray muon or electron neutrino interactions. We find that sMask-RCNN has an average pixel clustering efficiency of 85.9$\%$ compared to the dense network's average pixel clustering efficiency of 89.1$\%$. We demonstrate the ability of sMask-RCNN used in conjunction with MicroBooNE's state-of-the-art Wire-Cell cosmic tagger to veto events containing only cosmic ray muons. The addition of sMask-RCNN to the Wire-Cell cosmic tagger removes 70$\%$ of the remaining cosmic ray muon background events at the same electron neutrino event signal efficiency. This event veto can provide 99.7$\%$ rejection of cosmic ray-only background events while maintaining an electron neutrino event-level signal efficiency of 80.1$\%$. In addition to cosmic ray muon identification, sMask-RCNN could be used to extract features and identify different particle interaction types in other 3D-tracking detectors.}
\begin{document}
\maketitle
\flushbottom

\section{Introduction}
\label{sec:introduction}
The MicroBooNE \cite{ubooneDetector} experiment uses a liquid argon time projection chamber (LArTPC) with an active volume of 85 tonnes to study neutrinos from the Fermilab Booster Neutrino Beamline, while also receiving neutrinos from the Neutrinos at the Main Injector (NuMI) beam. The MicroBooNE LArTPC is a near-surface detector that does not utilize any overhead shielding for cosmic background mitigation. This, combined with a long TPC readout time, described in section \ref{sec:lartpcdesign}, results in a high ratio of cosmic ray muons to the number of neutrinos that interact within the detector. Techniques must be developed to reduce this cosmic ray muon background so that different neutrino interaction channels are measured with high purity.

\begin{figure}[htbp]
  \centering
    \begin{subfigure}{1.0\textwidth}
    \includegraphics[scale=.45]{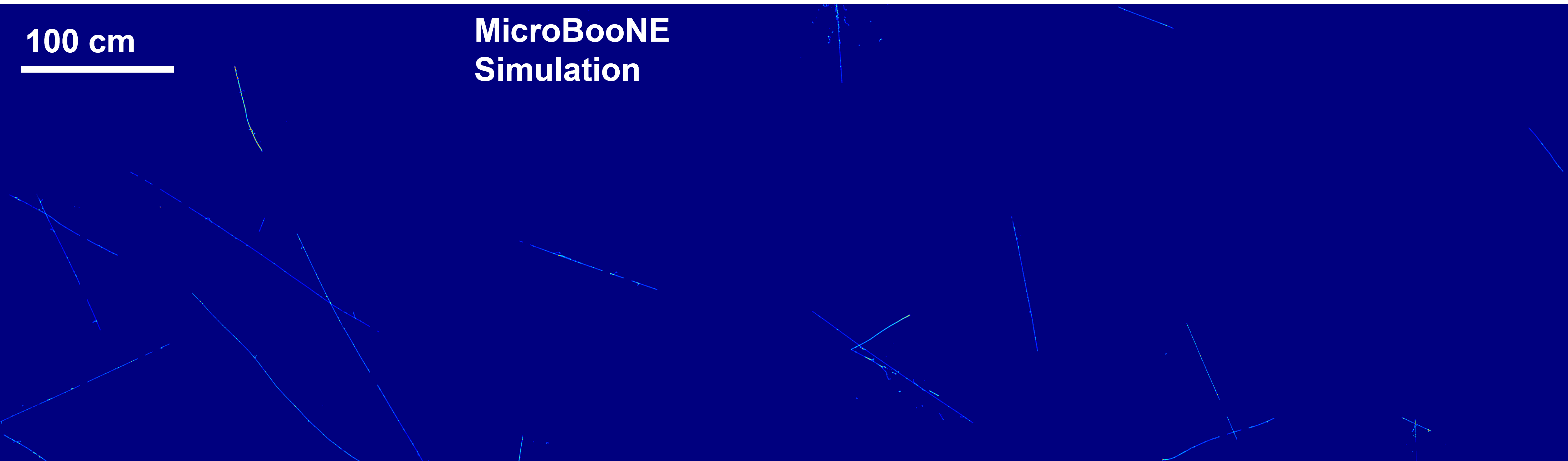}
    \caption{An example of an input image as given to sMask-RCNN to process}
    \label{fig:rawevd}
  \end{subfigure}
  \begin{subfigure}{1.0\textwidth}
    \includegraphics[scale=.45]{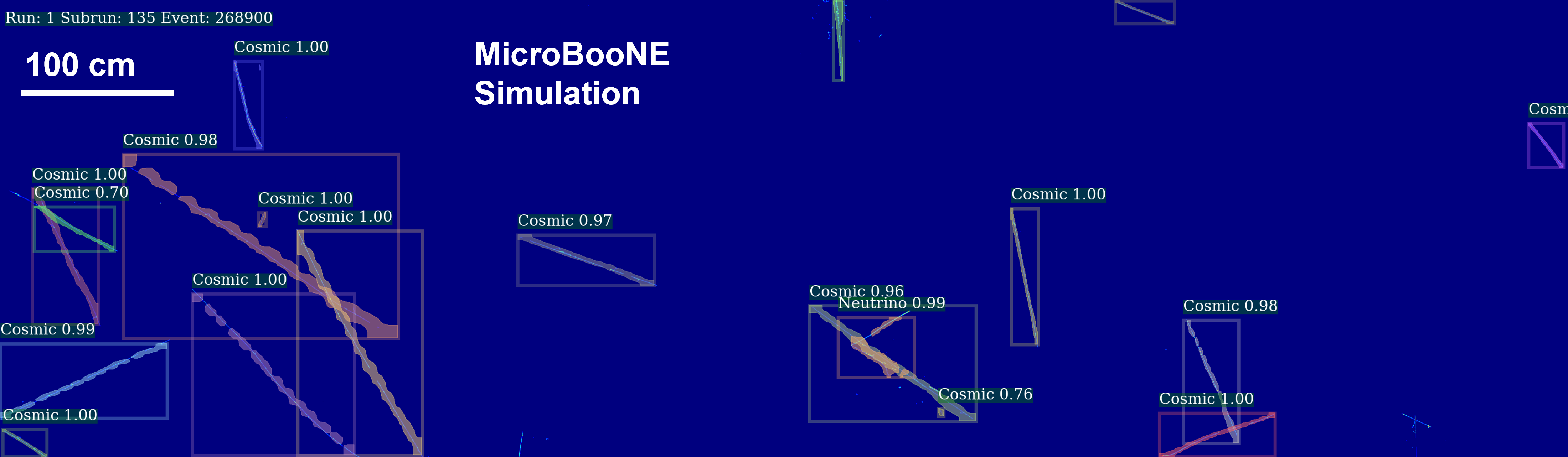}
    \caption{A simulated neutrino interaction overlaid on cosmic ray muons from data, labeled by sMask-RCNN}
    \label{fig:nuemc}
  \end{subfigure}
  \begin{subfigure}{1.0\textwidth}
    \includegraphics[scale=.45]{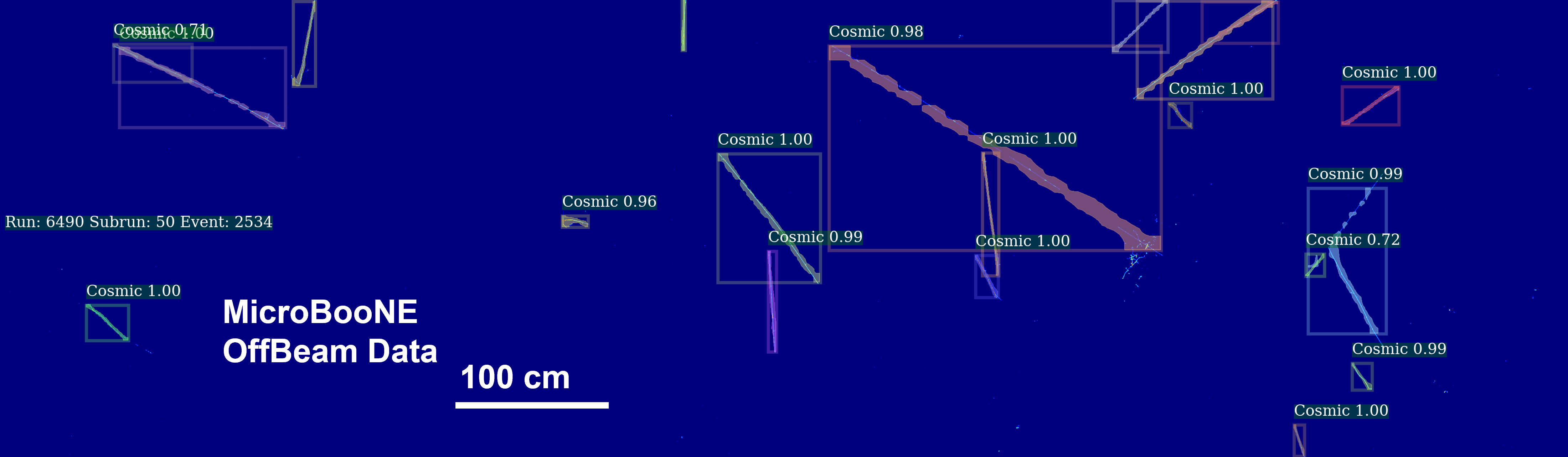}
    \caption{Cosmic-only data event, labeled by sMask-RCNN}
    \label{fig:extbnb}
  \end{subfigure}
  \begin{subfigure}{1.0\textwidth}
    \includegraphics[scale=.45]{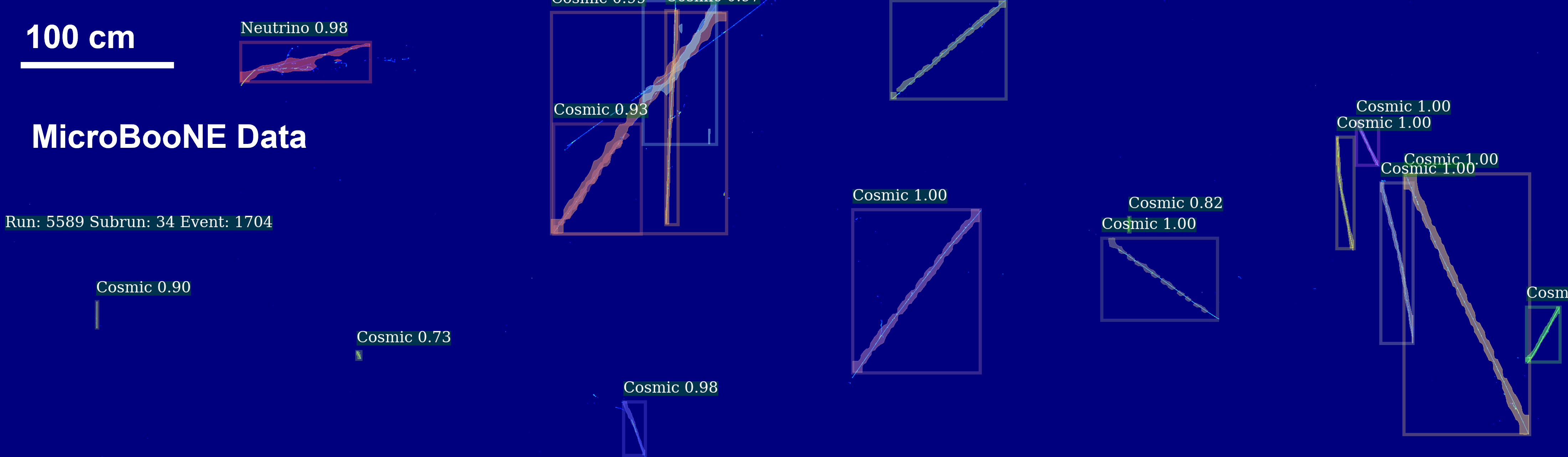}
    \caption{Data event containing a neutrino interaction, labeled by sMask-RCNN}
    \label{fig:nuedatasmrcnn}
  \end{subfigure}
  \caption{Several example event images. The vertical and horizontal scales are the same for all images. Each column of pixels along the $x$-axis refers to a specific wire readout, and each row along the $y$-axis refers to a different bin of signal readout time. This is described in greater detail in section \ref{sec:dataprep}. (a) is an example of an input image given to sMask-RCNN to process, whereas (b) shows the network’s subsequent labeling of the same image. (c) shows a cosmic-only data event. (d) shows a data event containing a neutrino interaction that sMask-RCNN correctly identifies with some confidence score, and clusters.}
  \label{fig:evdisps}
\end{figure}

A cosmic ray interaction can be mistaken for a neutrino interaction regardless of whether there is a neutrino interaction in the readout window. We consider any cosmic ray muon depositing charge in the detector an `interaction' regardless of whether it is captured in, decays in, or traverses the detector. Background events without a neutrino interaction present are called “cosmic-only” events. An example of a cosmic-only event is depicted in figure \ref{fig:extbnb}. Due to the prevalence of cosmic rays at this near-surface location, any reconstruction tools designed to tag cosmic ray muons should be deployed early in the reconstruction chain to filter such non-neutrino events. This means that our solution to identifying and removing cosmic ray backgrounds must be deployable across the entire MicroBooNE dataset. 

In this article, we present an approach to cosmic ray muon tagging using machine learning. Within the MicroBooNE experiment, machine learning techniques have been applied in other areas, such as particle identification, and pixel identification \cite{CONVNET,SSNET1,SSNET2,MPID}. In this article, we make use of a neural network called Mask-RCNN or "Mask Region-based Convolutional Neural Network" \cite{kaiminghe2017} to locate, identify, and cluster 2D interactions corresponding to the projections of the LArTPC wire planes. The design of Mask-RCNN is described in section \ref{sec:maskrcnnbackground}. 


Machine learning algorithms are typically deployed on graphics processing units (GPUs) because their ability to parallelize computations pairs well with the matrix multiplications that are abundant in machine learning code. However, in order to deploy on the full MicroBooNE dataset, our tools need to run on central processing units (CPUs) because the MicroBooNE production chain has access to large amounts of CPUs but not integrated GPUs. While running on GPUs would speed up run time, integrating GPUs would require additional personnel and financial investment that is not presently feasible, therefore operations that use CPUs are required. To solve this problem, we extend Mask-RCNN to use sparse submanifold convolutions \cite{DBLP:journals/corr/Graham14} which allow for much faster CPU running on sparse datasets by avoiding multiplication when one term is zero. We call this modified version of the network sparse Mask-RCNN or "sMask-RCNN". Section \ref{sec:submanifold} contains a brief description of submanifold convolutions, and sMask-RCNN's modified state is described in \ref{sec:sparsemrcnn}. Several visual examples of sMask-RCNN's performance are provided in the form of event images in figure \ref{fig:evdisps}. For details on how the event images underlying the sMask-RCNN labels are made, see section \ref{sec:dataprep}. Code detailing our implementation of sMask-RCNN has been made available at: \url{https://github.com/NuTufts/Detectron.pytorch/tree/larcv1_mcc9}.

\section{Background}
\label{sec:background}

\subsection{The MicroBooNE LArTPC}
\label{sec:lartpcdesign}
\begin{figure}[h]
\centering
\includegraphics[scale=.5]{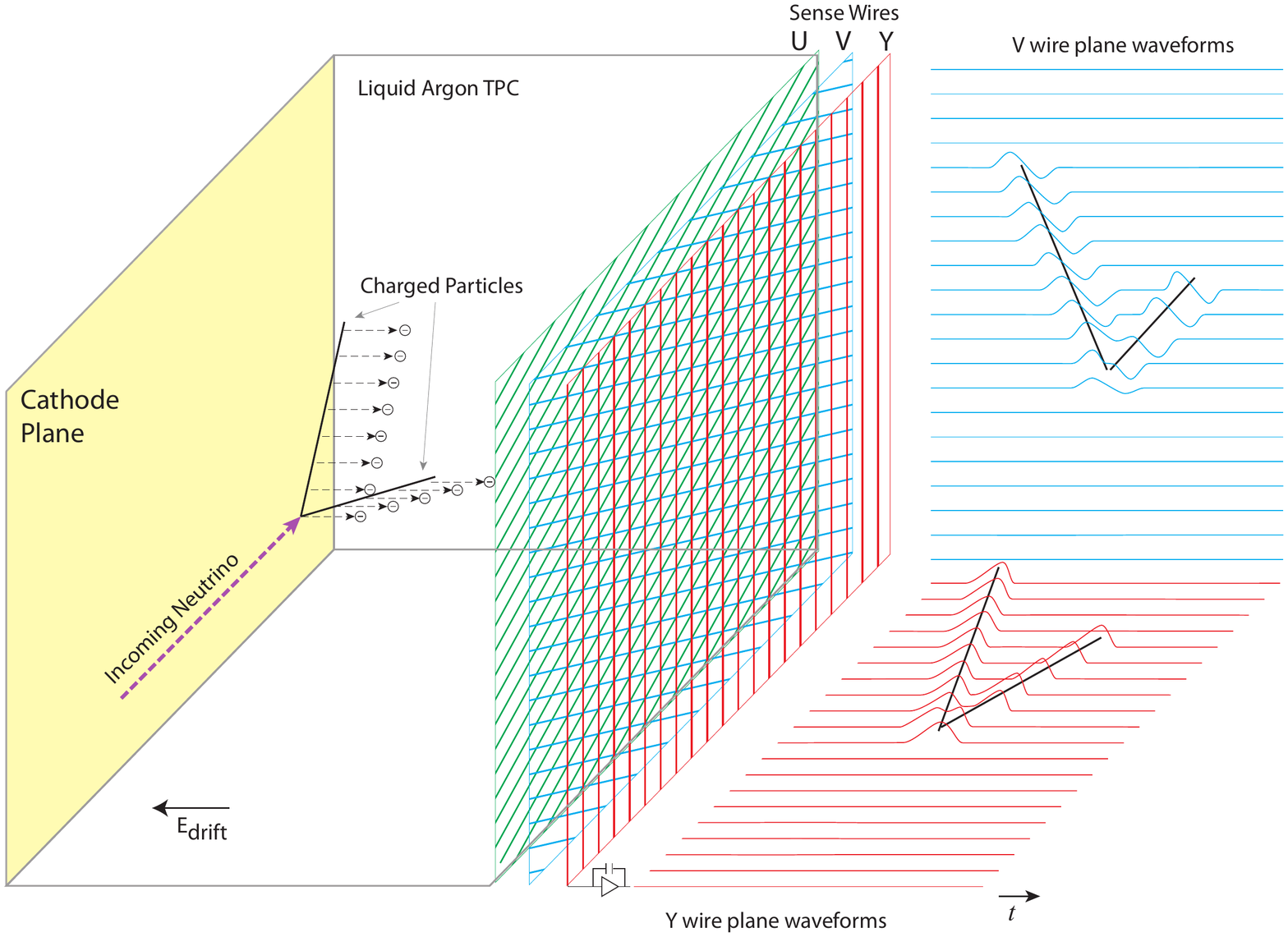}
\caption{A diagram of the LArTPC principle. The signal formation for the second induction plane (V plane) and the collection plane (Y plane) are shown \cite{ubooneDetector}.}
\label{fig:detdesign}
\end{figure}

The LArTPC technology is designed to provide precision calorimetry and particle tracking while remaining scalable to larger sizes for future experiments. In MicroBooNE’s LArTPC, a large volume of liquid argon is bounded on six sides within the time-projection chamber (TPC). On one side, the cathode, is a metal plate held at a negative potential of -70\,kV. On the other side of the argon, held near ground, is the anode: a collection of three wire planes at progressively higher potentials. Figure \ref{fig:detdesign} shows a diagram of the LArTPC principle.

Each wire plane consists of a series of parallel wires spaced every 0.3 cm. The wires in each plane are oriented at an angle of 60° with respect to the wires in the other two planes. When a charged particle passes through the detector, it creates ionization electrons which drift toward the anode wire planes due to the nearly uniform electric field between the cathode and anode. As the electrons pass by the first two wire planes they induce bipolar pulses on the wires before finally arriving at the third and final wire plane. Here they are collected and create a unipolar pulse. Thus this final wire plane is called the "collection plane". The bipolar and unipolar pulses read out from the wires undergo noise filtering and 2D deconvolution described in \cite{sigproc1,sigproc2}. This processing removes much of the noise from the wires and transforms the bipolar pulses into unipolar pulses. These wire signals are used to create the input images used by the neural network. The process for creating the input images is described in section \ref{sec:dataprep}.

\subsection{Existing cosmic identification tools}
\label{sec:existingtools}

Cosmic ray muon tagging and background removal have previously been performed in MicroBooNE using a variety of methods. One example uses deep learning with semantic segmentation to differentiate cosmic ray muon pixels from neutrino interaction pixels \cite{adamsmlremoval}. Mask-RCNN expands on semantic segmentation, further separating each instance of every individual interaction it finds using bounding boxes. This means that each cosmic muon interaction in the detector receives its own labeled and clustered output. 

Cosmic ray muon tagging has also been performed with more traditional algorithmic approaches. In the MicroBooNE experiment, one such method is the PandoraCosmic algorithm \cite{PANDORACOSMICPAPER1,PANDORACOSMICPAPER2}. This algorithm clusters hits in 2D and then combines these clusters into 3D tracks. It flags a track as a cosmic ray muon if part of the track is placed outside the detector based on timing information or if the track trajectory begins and ends at a TPC boundary using information related to the track's timing. It provides an exception for through-going trajectories that are parallel to the beam direction. This is designed to eliminate cosmic ray muons as they will appear as tracks originating from outside of the detector, and will be crossing perpendicular relative to the beam direction.

Another method used in MicroBooNE is the Wire-Cell (WC) cosmic tagger \cite{WCPHYSREV,MicroBooNE:2020jgj} which is made up of several event-level requirements, combined with the WC charge-light (Q-L) matching algorithm \cite{WCJINST}. This Q-L matching algorithm uses light information detected during the neutrino spill by 32 8-inch cryogenic photomultiplier tubes (PMTs) mounted behind the TPC wire planes. This light information is then spatially matched to charge deposited in the TPC, selecting TPC pulses created during the beam spill. Therefore, both the Q-L algorithm and the full WC cosmic tagger use additional information beyond the wire planes, which is what sMask-RCNN uses as input. In section \ref{sec:idbyneg} we show results achieved by a combination of sMask-RCNN and WC algorithms to produce a state-of-the-art cosmic ray tagger.

Cosmic ray tagging can also be achieved with hardware solutions. An example in MicroBooNE is the design and construction of the cosmic ray tagger system \cite{CRTPaper}. This system was introduced partway through MicroBooNE operation, and therefore is only available for part of the MicroBooNE data. It uses plastic scintillation modules to acquire the time and location for particles traversing the TPC. Reconstructed tracks can then be matched to this data and be flagged as pertaining to cosmic ray muons rather than neutrino interactions. Additionally, the cosmic ray tagger system can be used in tandem with software solutions to improve performance.

\subsection{Object detection and Mask-RCNN}
\label{sec:maskrcnnbackground}
The original Mask-RCNN network is designed to perform three common tasks in the field of computer vision: object detection, classification, and semantic segmentation. In the field of computer vision, classification is a task commonly performed to label an image as one of some predefined list of classes, for instance an image might be labeled a cat or a dog. Semantic segmentation refers to a labeling performed at pixel level, for example in an image with a cat and a dog, the pixels making up the dog are labeled `dog' pixels and those making up the cat are labeled `cat' pixels, while the remaining pixels are given a background label. The network is trained to receive some input image, place bounding boxes around objects of interest, classify these objects within some set of user defined classes, then within each box label each pixel as part of the object or not.

\begin{figure}[h]
\centering
\includegraphics[scale=.5]{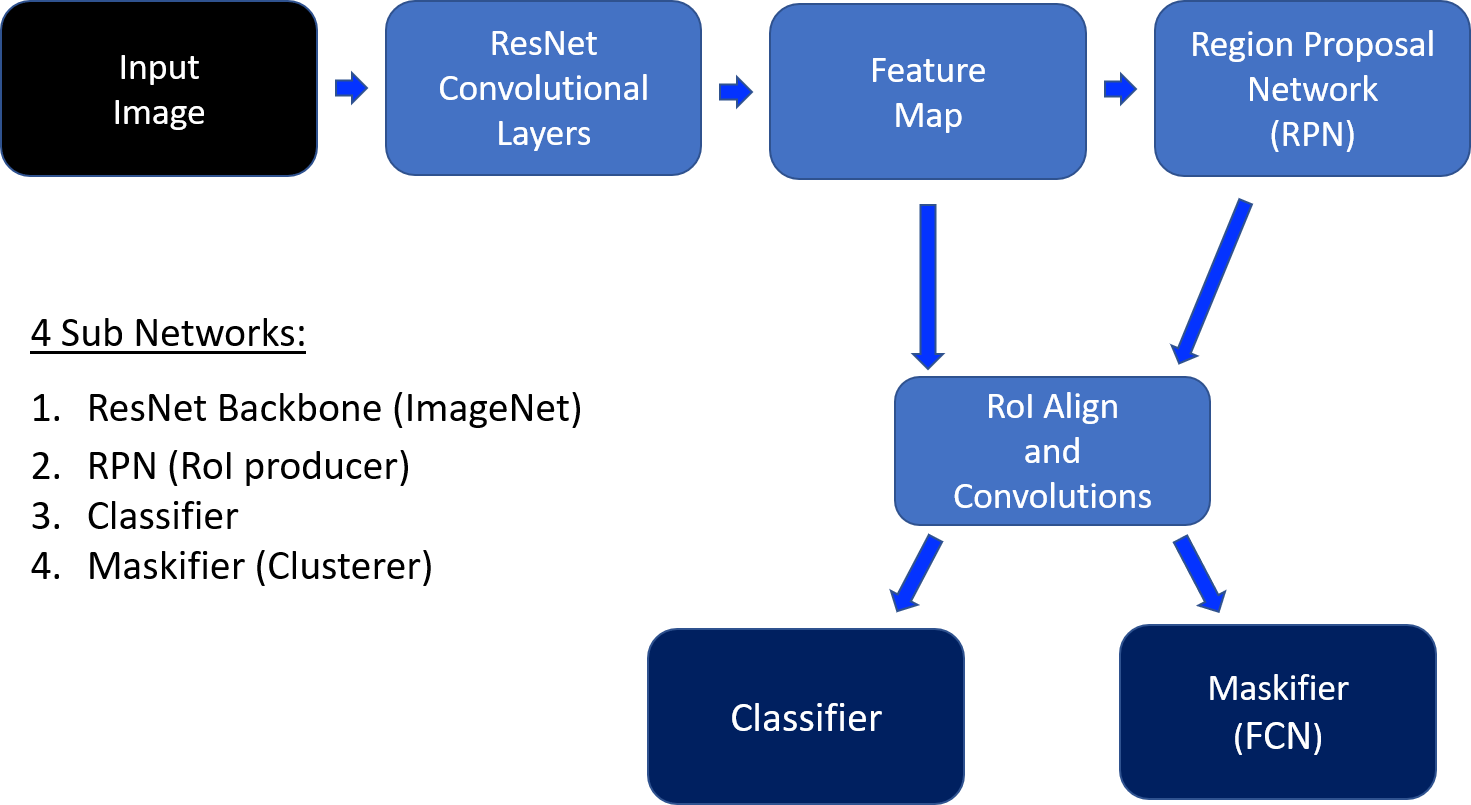}
\caption{Network Architecture for Mask-RCNN in MicroBooNE.}
\label{fig:netarch}
\end{figure}

Structurally, Mask-RCNN is comprised of four subnetworks. First is a residual network (ResNet) \cite{RESNETPAPER}. This network runs on the input image and creates a feature map for the image. This feature map is then fed into a region proposal network (RPN) \cite{ren2016faster}, which then produces a series of bounding boxes around regions of interest (RoIs) within the image. The bounding boxes are described by a 2D coordinate, a height and a width, and are designed to produce the smallest rectangular box containing the object. The RoIs are aligned in the feature map space via the RoIAlign algorithm, then combined with appropriate features, scaled to a fixed size and fed into the two final subnetworks: a classifier, and a fully convolutional network (FCN) we refer to as the "maskifier". The classifier takes each bounding box and its features and predicts which class of object it is with some confidence score. The maskifier produces a semantic segmentation mask of all the pixels within the box, determining which pixels correspond to the object and which are background. This semantic segmentation mask is synonymous with a cluster of pixels within the box, though the cluster need not be connected. Figure \ref{fig:netarch} shows a simplified view of the network architecture. 

\section{Methods}
\label{sec:methods}

\subsection{Data preparation}
\label{sec:dataprep}

To create images for analysis using MicroBooNE LArTPC data, we use the charge readout from the three wire planes at the anode. The neutrino beam window is 1.6\,ms, but we record a buffer on either side providing a modified window of about 3\,ms. The data taken during this 3\,ms comprises an "event". Over the course of this recorded beam window these wires are sampled as a rate of 2\,MHz. This equates to 6048 samples per wire. 

We create the event images shown in this article by placing the wire number along the $x$-axis and the sampling time along the $y$-axis. However, we first downsample the number of time samples per wire by a factor of six, going from 6048 samples per wire to 1008. This downsampling is performed both to reduce image size and to make the drift distance per pixel of an ionization electron roughly 0.3 cm. This distance matches the wire separation within a given wire plane. The value stored in each pixel within an image then  is proportional to the charge deposited in the detector. We will refer to this as the "pixel intensity". We then apply a threshold to the image by setting any pixel with pixel intensity below 10 (arbitrary units) equal to zero to further reduce noise. In comparison, the pixel intensity distribution from minimum ionizing particles peaks at $\sim 40$ in these arbitrary units.

While there is one image made for each of the three LArTPC wire planes, for this study we only use the collection plane. We choose the collection plane over the other wire planes because the collection plane does not require signal processing to turn bipolar pulses into unipolar pulses and therefore the signal is cleaner. The event images then have a dimensionality of the number of wires on the collection plane times the number of samplings per wire, or $3456 \times 1008$.

Lastly, in the MicroBooNE LArTPC a portion of the wires are unresponsive \cite{ubooneNoise}. In the collection plane this happens for about 10$\%$ of the wires. This creates an artifact in the images by creating vertical lines of unresponsive pixels. In some cases large groups of adjacent wires are unresponsive, leading to regions of unresponsive pixels.

\subsection{Sparse submanifold convolutions}
\label{sec:submanifold}

In the event images created from the LArTPC collection plane, about 0.7$\%$ of the pixels are nonzero, making the data "sparse". This is because the pixel value comes from the ionized electrons drifting away from the charged particles moving in the detector. Most of the time, the wires are reading out low-level noise that is below the threshold. The resultant low pixel occupancy means that when we apply Mask-RCNN to the event images, there are many computations that involve multiplications by zero. These trivial calculations waste computing resources, particularly if performed in sequence via a CPU and not in parallel via a GPU. 

\begin{figure}[h]
\centering
\includegraphics[scale=.40]{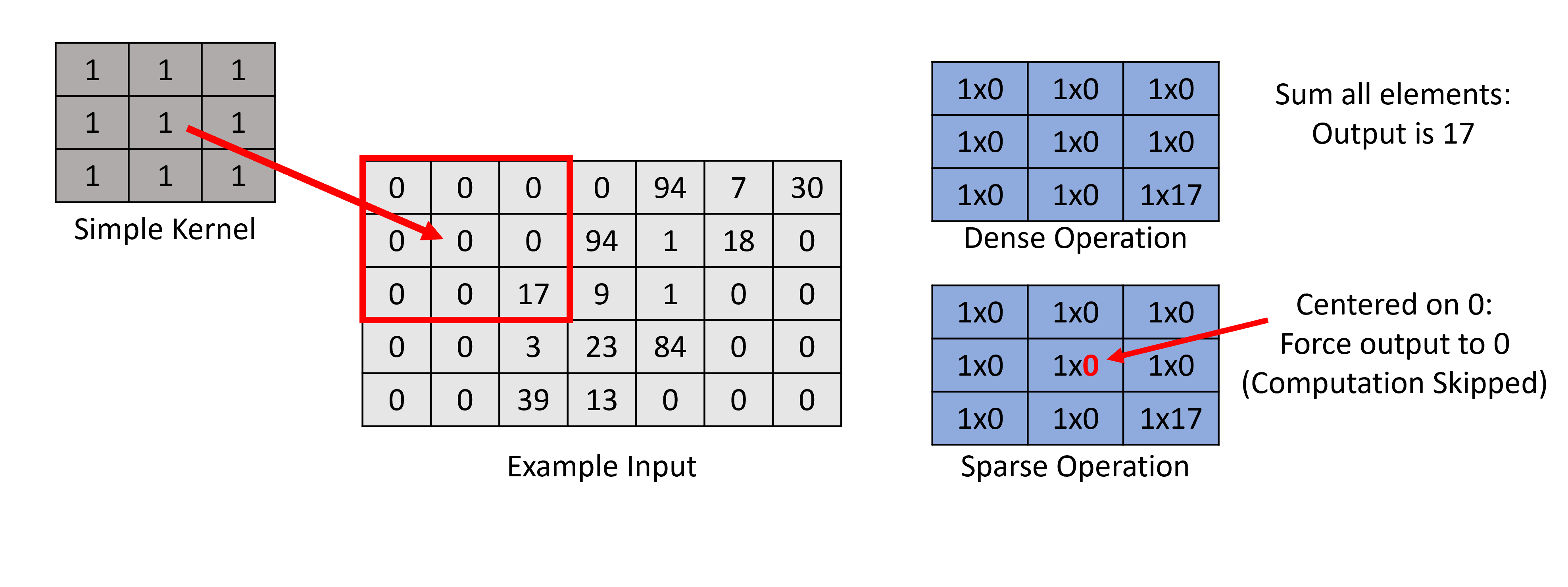}
\caption{An example of a convolution operation that depends on use of submanifold or normal convolutions. The normal convolution multiples the kernel against the image at the given position, and outputs a convolved value of 17. Meanwhile, the submanifold convolution does not get computed when centered on a zero. The submanifold convolutions used in sMask-RCNN have a kernel size of $3 \times 3$ (same as shown in the figure), with a stride of 1.}
\label{fig:convolutionexample}
\end{figure}

A normal convolution in a neural network takes a convolutional kernel or filter and moves it across the image, multiplying at each location to acquire the convolved value. In comparison, a sparse submanifold convolution only multiplies the kernel against positions centered on nonzero pixels, avoiding computations on zeroed regions of the input. Notably, a submanifold convolution is not mathematically equivalent to a regular convolution. In normal convolutions, kernels centered on zeros in the input image can output a nonzero convolved value if the edge of the filter captures some nonzero input. An example of this is shown in figure \ref{fig:convolutionexample}. This blurs the features coming out of a convolution, so that convolutions which are performed one after another in a deep neural network, such as ResNet, spread information outward. 

\subsection{Sparse Mask-RCNN}
\label{sec:sparsemrcnn}

We utilize Mask-RCNN to locate, identify, and cluster interactions within the 2D event images described in section \ref{sec:dataprep}. The network places bounding boxes around, classifies, and then clusters pixels corresponding to deposited charge for each interaction it finds within the image. We define an interaction as all of the charge deposited in an image coming from the same ancestor particle. For example, if an electron neutrino interacts with the argon in the LArTPC, and yields a proton and an electron, all of the charge deposited from the proton and electron are combined into one ancestor "electron neutrino" interaction and should be clustered together. A more detailed description of network training is described in section \ref{sec:training}.

To speed up the network when deployed on CPUs, we swap the ResNet convolutions with sparse submanifold convolutions while maintaining the original network structure.  Additional work could be done to make the later subnetworks use sparse convolutions but examinations of the compute times for the individual parts of sMask-RCNN made this unnecessary. For clarity, we will refer to Mask-RCNN without submanifold convolutions as dense Mask-RCNN.

The change to sparse convolutions yields a significant gain in terms of network speed for inference when running on CPUs. The timing information for running the dense and sparse configurations of the network is shown in table \ref{tab:time_table}. These timings were performed on an Intel(R) Core(TM) i9-9820X CPU @ 3.30 GHz and measure wall time. This allows the network to be inserted into production code on CPU farms such as FermiGrid \cite{FERMIGRID} and to scale up how quickly the network is run over large data samples. We note that the Intel CPU we tested on is superior to what is generally available on CPU farms. Brief testing done on available CPU farms shows the ResNet runtime difference is exacerbated. This means the difference between sparse and dense implementations is even greater when older CPUs are used. This further prioritizes shortening the ResNet runtime, as when we deploy on CPU farms, we will use a variety of CPUs with lesser performance than an Intel(R) Core(TM) i9-9820X CPU @ 3.30 GHz. Further, an added benefit of using CPUs is that this technique is scalable to future experiments and studies where more data may be analyzed, and the computing resources cannot scale to a reasonable number of GPUs. 

The implementation to sparse ResNet also introduces improvements to training when it comes to memory. For the dense version of the network, due to memory constraints, we could only train the network on $832 \times 512$ crops of the event image. We use the word crop to refer to a random cutout of the original image of this new $832 \times 512$ size, where the crop must contain a portion of a simulated interaction. However, sMask-RCNN is trained on the full $3456 \times 1008$ event images because the memory required on the GPU to store a full image for the network is reduced by roughly a factor of the image's occupancy, as all zero pixels are no longer operated on. Regardless of training size, both the dense and sparse forms of Mask-RCNN could be deployed on the full $3456 \times 1008$ event images, as less memory is used if not actively training the network.

\begin{table}
\centering
\caption{\label{tab:time_table}The average inference runtimes per $3456 \times 1008$ pixel image on a CPU. The first row is the runtime for just the ResNet portion of Mask-RCNN on the images. The second row is the time to run the entire network on the images. In the case of sparse ResNet, the time spent making the input image into a sparse tensor and the output features into a dense tensor is included in the sparse ResNet module time.}
\begin{tabular}{||c  c  c||} 
\hline
 & Dense ResNet & Sparse ResNet \\ [0.5ex] 
\hline\hline
ResNet Runtime &   3.172 s   & 0.1758 s  \\ 
\hline
Full Detection Runtime &   8.438 s   & 5.79 s \\
\hline
\end{tabular}
\end{table}

All of the event images shown in this article use the sparse implementation of Mask-RCNN. We examine the difference in performance between the dense and sparse networks in section \ref{sec:results}.

\subsection{Network training}
\label{sec:training}

When training, the entire network performs a forward pass on an image, and then the backward pass updates the weights of all the subnetworks based on a combined loss function built from the outputs of the maskifier, classifier, and RPN, as described in the original Mask-RCNN article \cite{kaiminghe2017}. Rather than train the ResNet from randomized initial weights, we use weights pretrained on the ImageNet dataset \cite{IMAGENET}, which is a publicly available labeled dataset of images of animals and everyday objects. The ImageNet dataset is commonly used in the field of computer vision. We briefly started training ResNet from scratch, with randomly initialized weights, because the ImageNet pretraining is designed to identify animals and everyday objects, but despite this, we found that using the pretrained ResNet gave more useful features for the other components of Mask-RCNN to utilize in this particle physics analysis.

To train both the dense and sparse forms of Mask-RCNN, we use a sample of simulated electron neutrino events featuring simulated cosmic background. This means that every full $3456 \times 1008$  image contains a single electron neutrino interaction among many cosmic ray muons. While the dense network was trained on crops containing at least part of a neutrino interaction or cosmic ray muon, they did not always have an example of both within the same crop.

The interactions present in the training data are broken up into six different interaction classes, detailed in table \ref{tab:class_table}. Dense Mask-RCNN was trained on each of these interactions, but as we determined our goal was primarily cosmic ray muon tagging, we only trained the sparse implementation on cosmic ray muons and electron neutrino interactions. This is because our implementation of this network is designed to run at the beginning of a reconstruction chain, to tag cosmic-ray muons at the start, remaining backgrounds can be dealt with using more targeted approaches downstream. Simulated interactions where the simulated ancestor particle was one of the four other classes were still present in the data but the network was told to ignore them. The training data uses CORSIKA \cite{CORSIKA} for cosmic ray muon simulation. For electron neutrino interactions the GENIE neutrino interaction simulator \cite{GENIE} is used. In both instances, GEANT4 \cite{GEANT4} is used to model the detector response.

We also note that the version of simulation used to produce the training data for the sparse network is slightly updated as a newer version of simulation became available. Both the dense and sparse networks have their performance evaluated on the same set of 9400 events from the newer version of simulation, meaning there is a slight discrepancy between the training and testing data for the dense network. However, we expect this change to have minimal effect on the network performance.

\begin{table}[h]
\centering
\caption{\label{tab:class_table}The different class types and number of occurrences in the training sets for the dense and sparse versions of Mask-RCNN. Note that the sparse network only trained on cosmic ray muon and electron neutrino interactions.}
\begin{tabular}{||c | c | c || c | c||}
\multicolumn{1}{c}{} & \multicolumn{2}{c}{Dense Training} & \multicolumn{2}{c}{Sparse Training} \\
\hline
Interaction Class & Counts & Percentage & Counts & Percentage \\ [0.5ex] 
\hline\hline
Cosmic Ray Muon &   2708730   & 92.99 &   786050   & 95.24  \\ 
\hline
Electron Neutrino &   97034   & 3.33 &   39296   & 4.76 \\
\hline
Neutron &   26072   & 0.90 & - & - \\
\hline
Proton &   5738   & 0.19 & - & - \\
\hline
Electron &   155   & 0.005 & - & - \\ 
\hline
Other &   75026   & 2.58 & - & - \\ [1ex] 
\hline
\end{tabular}
\end{table}

As is common practice in machine learning, we split the data into two orthogonal subsets: a training set with 80$\%$ of the events and a validation set with 20$\%$. When training the network, we use events in the training set, and whenever we wish to measure the performance of the network, such as calculating performance metrics shown in section \ref{sec:results}, we use the validation set. This is a critical part of machine learning because it verifies that the network can generalize and perform its task on events outside of the training set.  

The dense version of the network is trained on a sample of 230,000 crops for 1.75 epochs. In the context of machine learning, an epoch is one training pass through the data. While 1.75 is a low number of epochs, each crop features multiple interactions seen and masked by the network. Then the network is trained on a subset of 30,000 of these crops containing examples of high intersection-over-union (IOU) interactions. IOU between two interactions is defined as the number of pixels present in both interaction bounding boxes, divided by the total number of unique pixels present in either bounding box. These crops featured multiple interactions with overlapping bounding boxes. This fine tuned training is performed due to poor performance by the dense network on overlapping interactions. The training on this subset is performed for 8 epochs so the network can focus on learning these difficult events. 

The sparse version of the network is trained on a sample of 40,000 full event images for three epochs. No fine tuning needed to be performed on highly overlapping interactions as the sparse network did not appear to suffer from the same issue as the dense network. It should be noted that while at first glance the training sample sizes of the dense and sparse networks differ by a significant factor, in actuality the dense crops were made at a factor of up to 10 crops per full image, and the training datasets are comparable in terms of interaction sample variance. Further, all performance evaluations for both networks are deployed on the same validation set of 9400 full sized images. 

\section{Comparing dense and sparse performance}
\label{sec:results}

In this section we define several metrics to test the performance of Mask-RCNN at identifying and clustering interactions within MicroBooNE event images. Then we compare the performance between the dense and sparse versions of the network. For this evaluation, we include interactions found by the network with a class score of 0.4 or higher. The class score is a score between 0 and 1.0 indicating how confident the network is that the class label is correct. This threshold is chosen to provide a balance between the purity and efficiency metrics defined and discussed below. Once this threshold is applied, all remaining predicted interactions are treated equally for the purpose of calculating metrics.

All analysis in this section uses the full scale $3456 \times 1008$ event images in the validation set of simulated data used to train the sparse network. This refers to the data described in the right two columns of table \ref{tab:class_table}.

\subsection{Efficiency and purity}
\label{sec:effpur}

\begin{figure}[h]
\centering
\includegraphics[scale=.40]{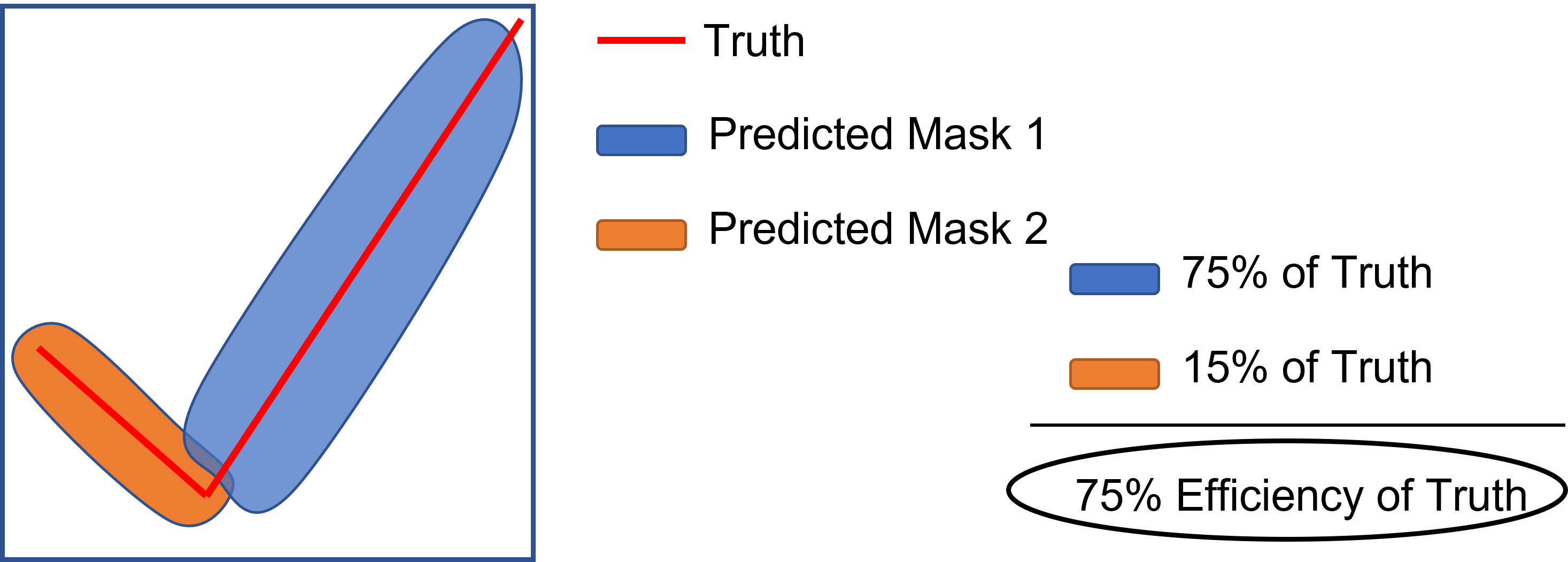}\caption{A visual representation of the definition of efficiency. Only nonzero pixels in the event image are counted.}
\label{fig:eff_def}
\end{figure}

\begin{figure}[h]
\centering
\includegraphics[scale=.40]{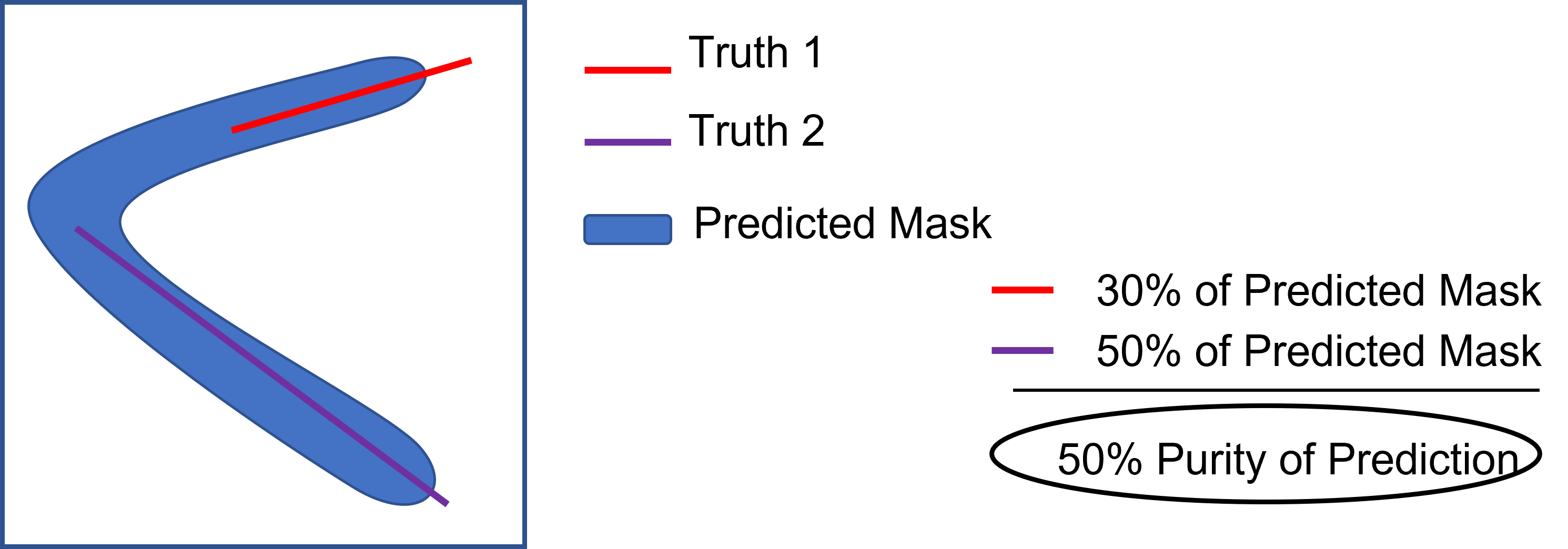}\caption{A visual representation of the definition of purity. Only nonzero pixels in the event image are counted.}
\label{fig:pur_def}
\end{figure}

For the purpose of this section, we will refer to interactions either as "true" or "predicted". A true interaction is one that exists in the simulation, whereas a predicted interaction is one that the network claimed it has found and labeled. We define two metrics to measure the pixel level efficiency and purity of the network's ability to find and cluster interactions. The efficiency is a measure of the percentage of pixels in a true interaction that are masked by the network's prediction. Purity is defined for each predicted interaction as the highest fraction of pixels belonging to the predicted interaction and a single true interaction. For example, if 30$\%$ of a prediction maps to true interaction A, and 50$\%$ maps to true interaction B then the purity is 50$\%$ for that predicted interaction. In both of these definitions, only pixels containing charge deposition above the pixel intensity threshold of 10 are considered, as we do not care about clustering empty pixels. Concretely, the efficiency $E$ is defined as 
\begin{equation}
\label{eqn:eff}
E = \frac{\sum_{ij} T_{ij} \cdot W_{ij} \cdot M_{ij} } {\sum_{ij} T_{ij}\cdot W_{ij}},
\end{equation}
where $T$, $W$, and $M$ are matrices representing the truth interaction, wire event image, and predicted interaction mask, respectively, with dimensions of the event image. Meanwhile, $i$ and $j$ are pixels indices. A visual representation of the efficiency calculation is depicted in figure \ref{fig:eff_def}. Similarly, the purity $P$ is defined as 
\begin{equation}
\label{eqn:pur}
P = \frac{\sum_{ij} M_{ij} \cdot W_{ij} \cdot T_{ij} } {\sum_{ij} M_{ij}\cdot W_{ij}}.
\end{equation}
The purity calculation is depicted in figure \ref{fig:pur_def}. For each of these equations, the values in $T$ are 1 if the pixel belongs to the true interaction, and 0 otherwise, while the values in $M$ are 1 if the pixel belongs to the predicted interaction mask, and 0 otherwise. Finally, the values in $W$ are 1 if the pixel has any deposited charge, and 0 otherwise. After the element-wise multiplication of the matrices, the summations then run over the indices of the matrix. This corresponds to a counting of the pixels corresponding to the union of the given matrices. 

A true interaction's efficiency is taken as the best value as calculated for all predicted interactions. A predicted interaction's purity is taken as the best value when calculated for all true interactions. 

These definitions mean that a given event image will have one efficiency measurement for each true interaction, and one purity measurement for each predicted interaction. While we are aware that object identification customarily uses panoptic quality \cite{panopticqual} or intersection-over-union as evaluation metrics, we choose to use efficiency and purity in better keeping with particle physics analysis language.

\begin{figure}
\centering
\begin{subfigure}{.5\textwidth}
  \centering
  \includegraphics[scale=0.5, width=\linewidth]{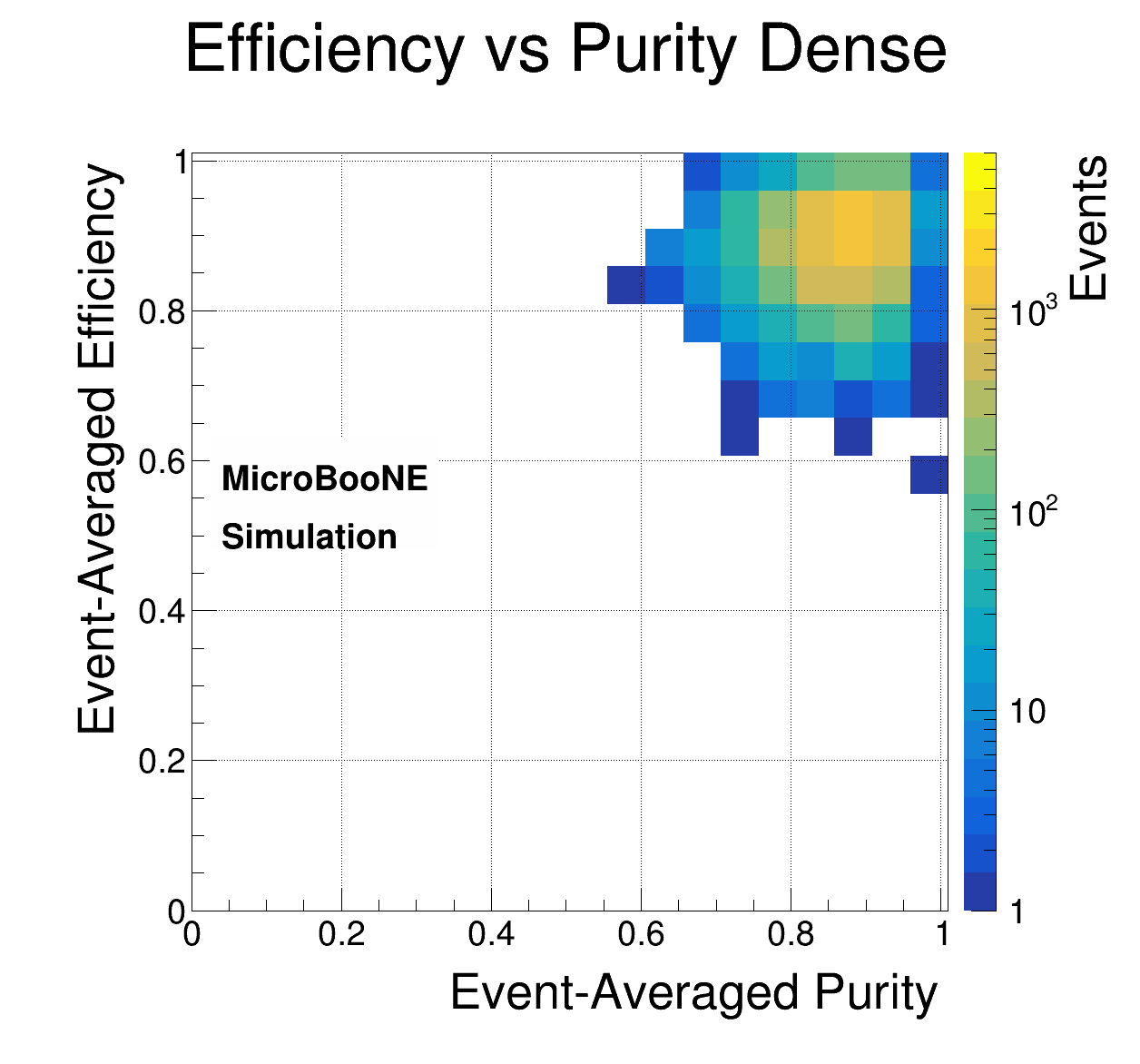}
  \caption{Dense Mask-RCNN}
  \label{fig:dscompare_2dEffPur1}
\end{subfigure}%
\begin{subfigure}{.5\textwidth}
  \centering
  \includegraphics[scale=0.5, width=\linewidth]{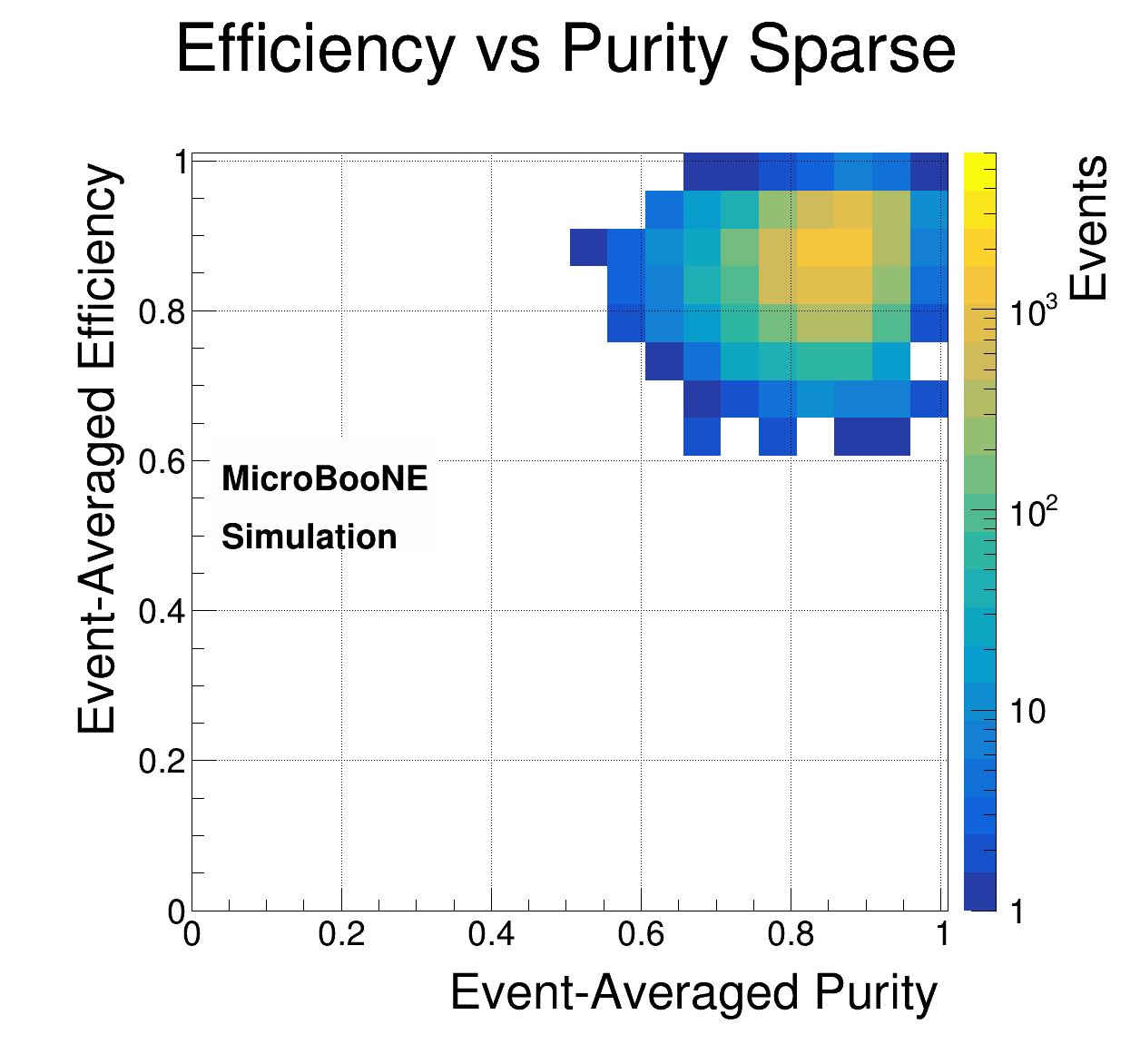}
  \caption{Sparse Mask-RCNN}
  \label{fig:dscompare_2dEffPur2}
\end{subfigure}
\caption{The event-averaged efficiencies and purities for the dense and sparse implementations of Mask-RCNN. The dense network has a mean event-averaged efficiency of 0.89 and a mean event-averaged purity of 0.87. For the sparse network these values are 0.86 and 0.85. Each of these evaluations use the same validation dataset.}
\label{fig:dscompare_2dEffPur}
\end{figure}

For each event, we average the purities and efficiencies for the predicted and true interactions. These averages are of O(20) interactions, where there is a single neutrino interaction and many cosmic ray muon interactions. We reiterate that a cosmic ray muon `interaction' is just any cosmic ray muon and potential daughter particles that deposit charge in the detector. These event-averaged values are plotted in the 2D histograms shown in figure \ref{fig:dscompare_2dEffPur}.

\begin{figure}
\centering
\begin{subfigure}{.5\textwidth}
  \centering
  \includegraphics[scale=0.5, width=\linewidth]{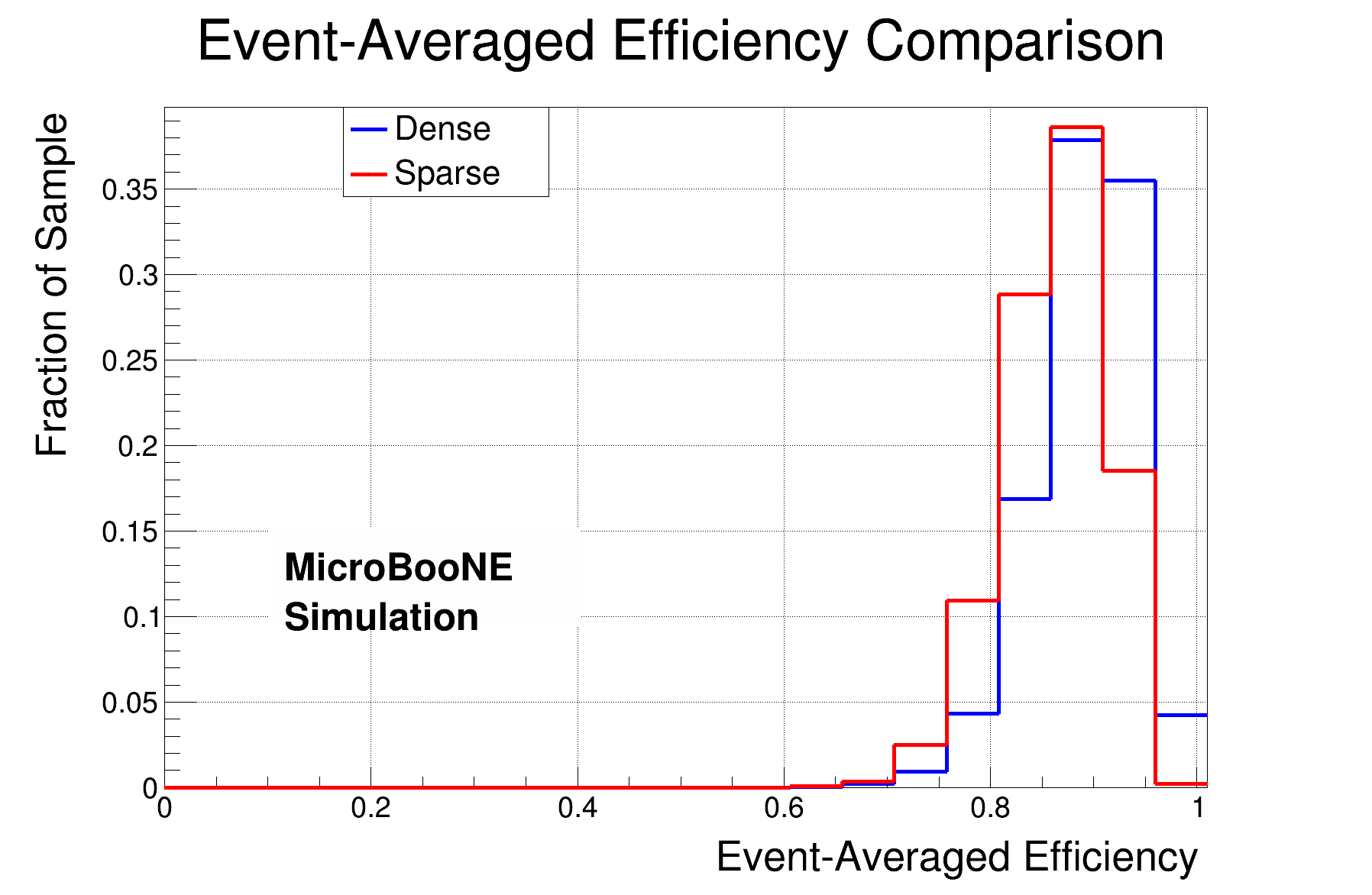}
  \caption{Event-Averaged Efficiency}
  \label{fig:ev_avg_eff}
\end{subfigure}%
\begin{subfigure}{.5\textwidth}
  \centering
  \includegraphics[scale=0.5, width=\linewidth]{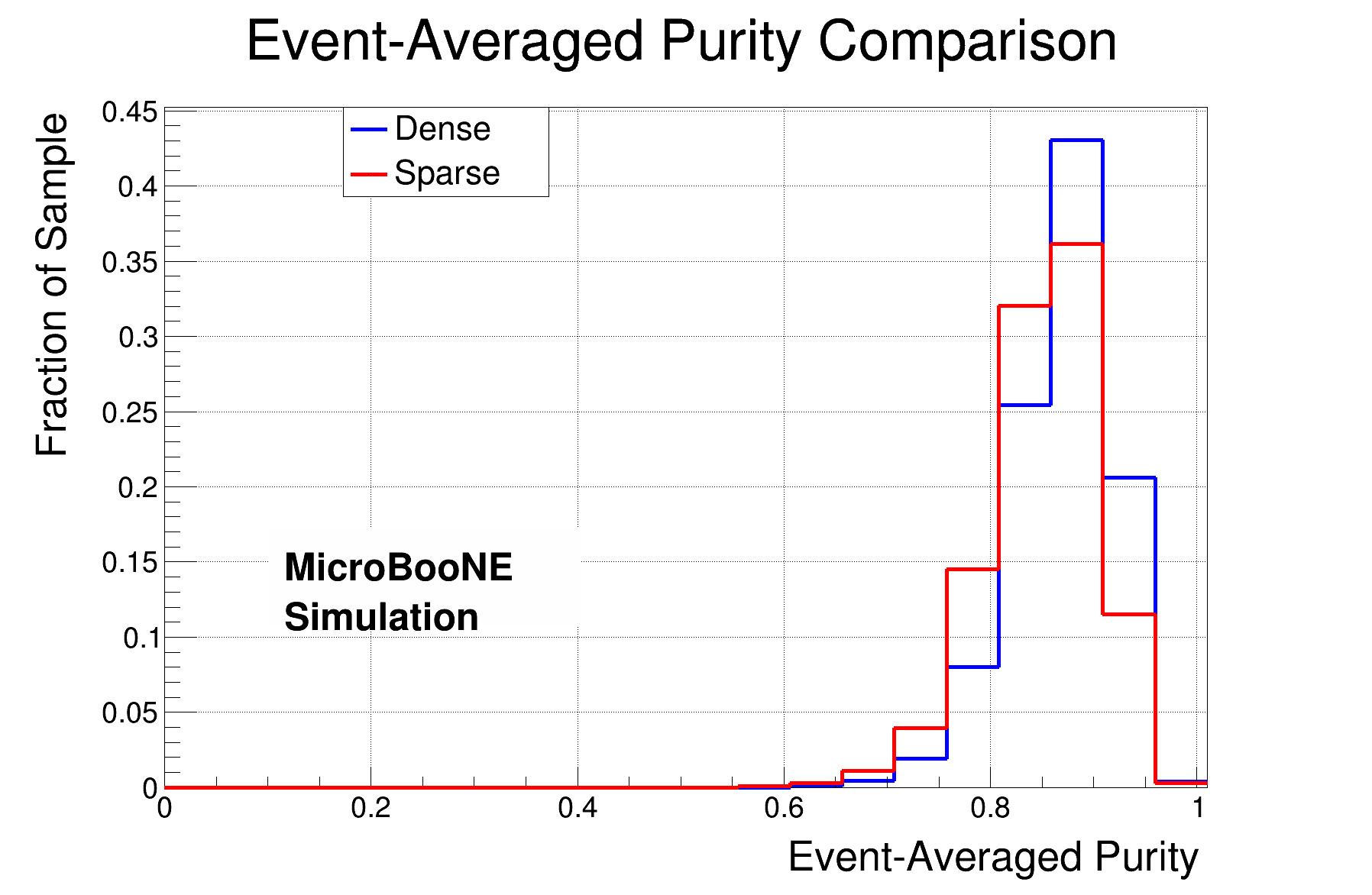}
  \caption{Event-Averaged Purity}
  \label{fig:ev_avg_pur}
\end{subfigure}
\caption{The one-dimensional distributions for the event-averaged efficiency (a) and purity (b) shown in figure \ref{fig:dscompare_2dEffPur}. Each plot compares the dense and sparse network performances. }
\label{fig:ev_avg_projections}
\end{figure}

Perfect efficiency and purity would yield values of 1.0 for each, so these histograms have targets in the upper right corners. We can see that the event-averaged purity drops from 87$\%$ for the dense to 85$\%$ for the sparse, while the event-averaged efficiency drops from 89$\%$ to 86$\%$. The one-dimensional projections of figure \ref{fig:dscompare_2dEffPur} are shown in figure \ref{fig:ev_avg_projections}, where the event-averaged efficiency and purity distributions are compared between the dense and sparse networks.

\begin{figure}[h]
\centering
\includegraphics[scale=.2]{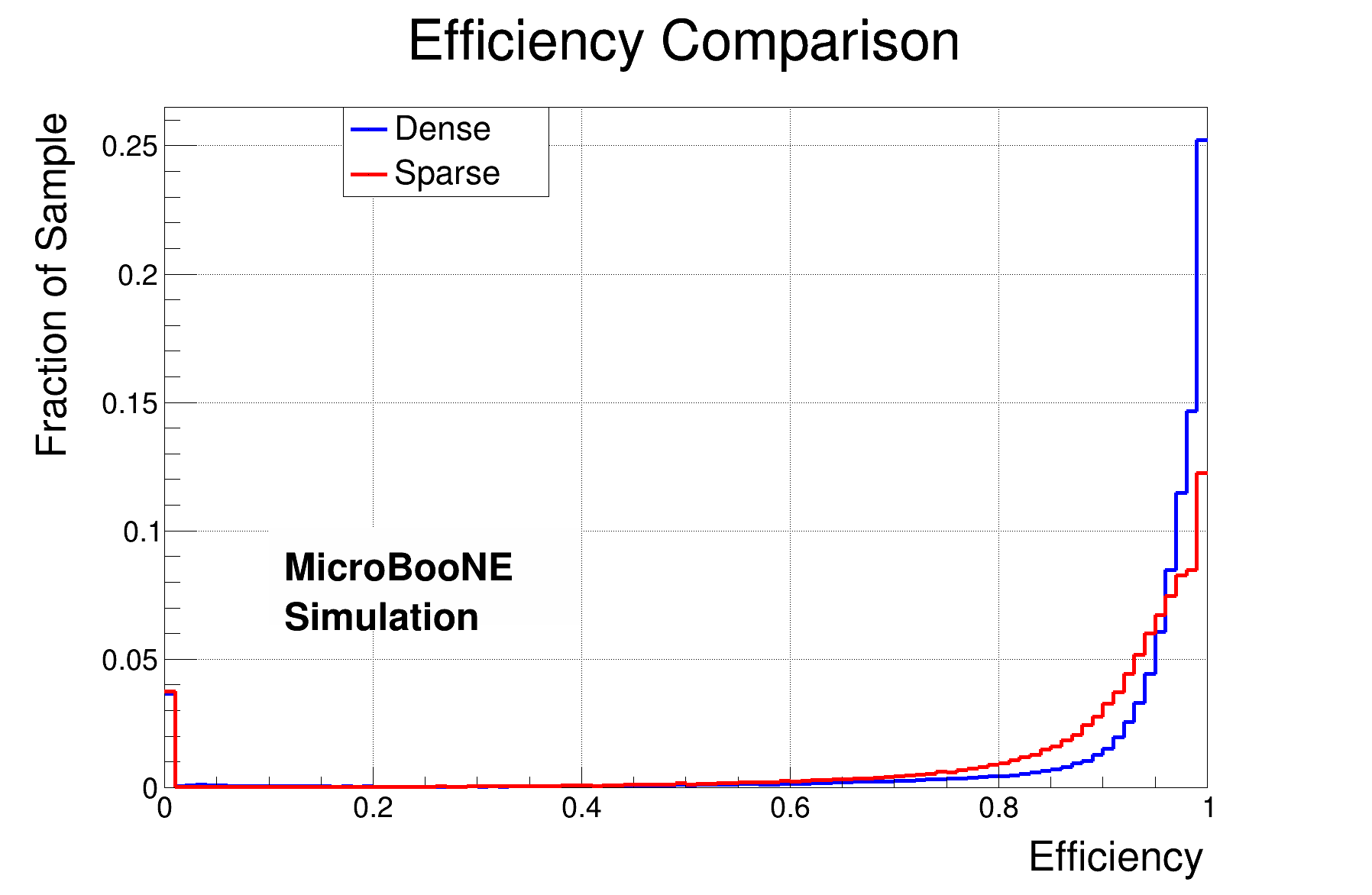}
\caption{The interaction-level efficiency of the dense and sparse versions of Mask-RCNN as measured on the validation set.}
\label{fig:efficiency_overlaid}
\end{figure}

It is also useful to examine the individual interaction efficiencies, rather than the event-averaged versions. The distributions for both the dense and sparse versions of Mask-RCNN are shown in figure \ref{fig:efficiency_overlaid}. Here we can see the sparse (red) distribution is worse than the dense (blue) distribution. Notably the size of the peak at zero is the same for the two versions of the network. A true interaction will have zero efficiency if the network has no prediction that masks part of it. The fact that the two versions have the same sized peak at zero indicates that they each find the same number of interactions, but the dense masks are somewhat more complete.

\begin{figure}[h]
\centering
\includegraphics[scale=.2]{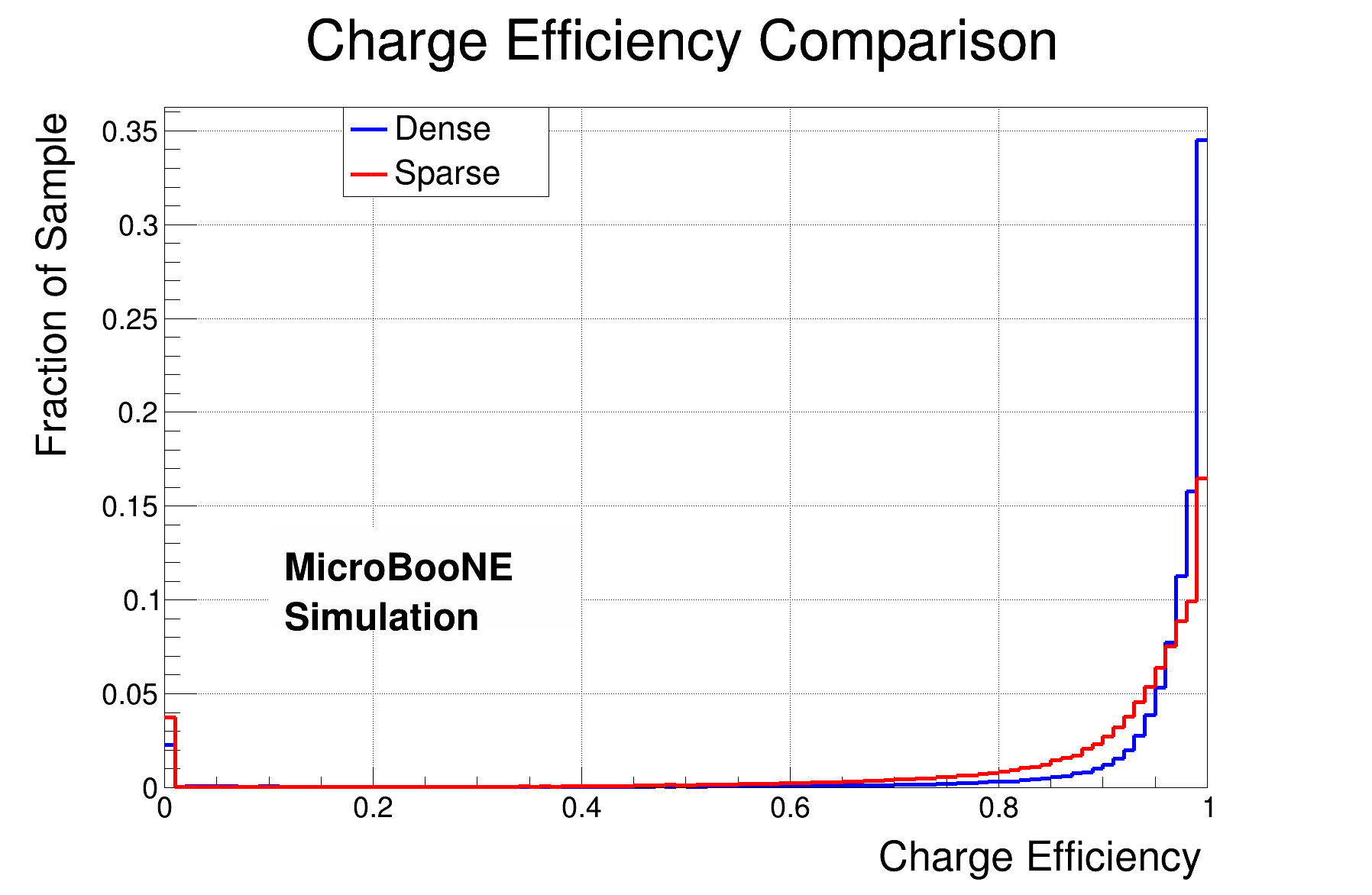}
\caption{The interaction-level charge efficiency of the dense and sparse versions of Mask-RCNN as measured on the validation set.}
\label{fig:charge_eff_overlaid}
\end{figure}

The efficiency calculation is modified slightly by weighting the pixels within an interaction by their deposited charge. This version of the efficiency we term "charge efficiency" and is shown in figure \ref{fig:charge_eff_overlaid}. Here we see a shift to the right for each distribution compared to their pixel-level efficiency in figure \ref{fig:efficiency_overlaid}. This indicates the network's preference for clustering higher value pixels corresponding to larger deposited charge, though this result may be due to the network being more likely to grab the center of tracks in our image, where the higher value pixels lie, compared to the halo of hits along a track's edge. While not surprising, this is a useful feature as the physics quantity we are dealing with is charge, not pixel count.

\begin{figure}[h!]
\centering
\includegraphics[scale=.45]{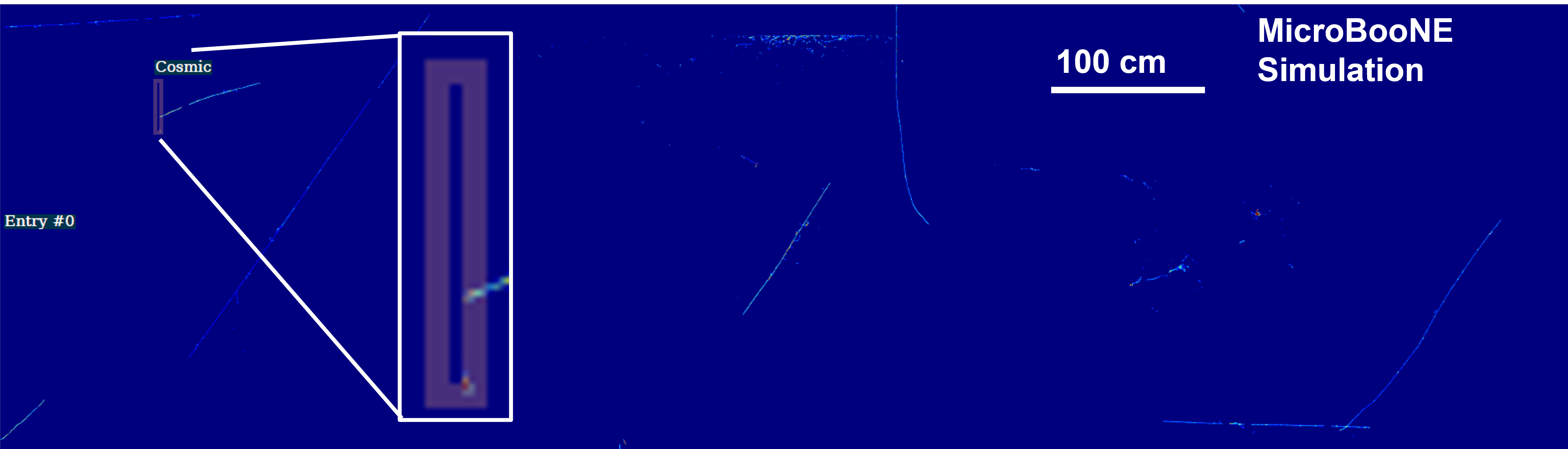}
\caption{A zero-efficiency true interaction almost entirely in a region of unresponsive wires. The white box shows a zoom-in of the area of interest, and within it, the colored box should contain true neutrino interaction. However, because this interaction falls in an unresponsive region of the detector, no deposited charge is seen inside the colored box.}
\label{fig:ZeroEff_Dead}
\end{figure}

If we explore the interactions lying within the zero efficiency peaks in these plots then we find two common failure modes. The first is made up of interactions that lie completely or significantly in the unresponsive regions of the event image. Recall that roughly 10$\%$ of the MicroBooNE LArTPC wires are unresponsive, corresponding to vertical lines of unresponsive regions in the event images. True interactions within these regions are in the simulation, but have little in the way of signal in the event image for the network to detect. An example of a true interaction simulated in an unresponsive region of the image is shown in figure \ref{fig:ZeroEff_Dead}. It is unreasonable to expect the network to be able to label such interactions.

\begin{figure}[h]
\centering
\includegraphics[scale=.45]{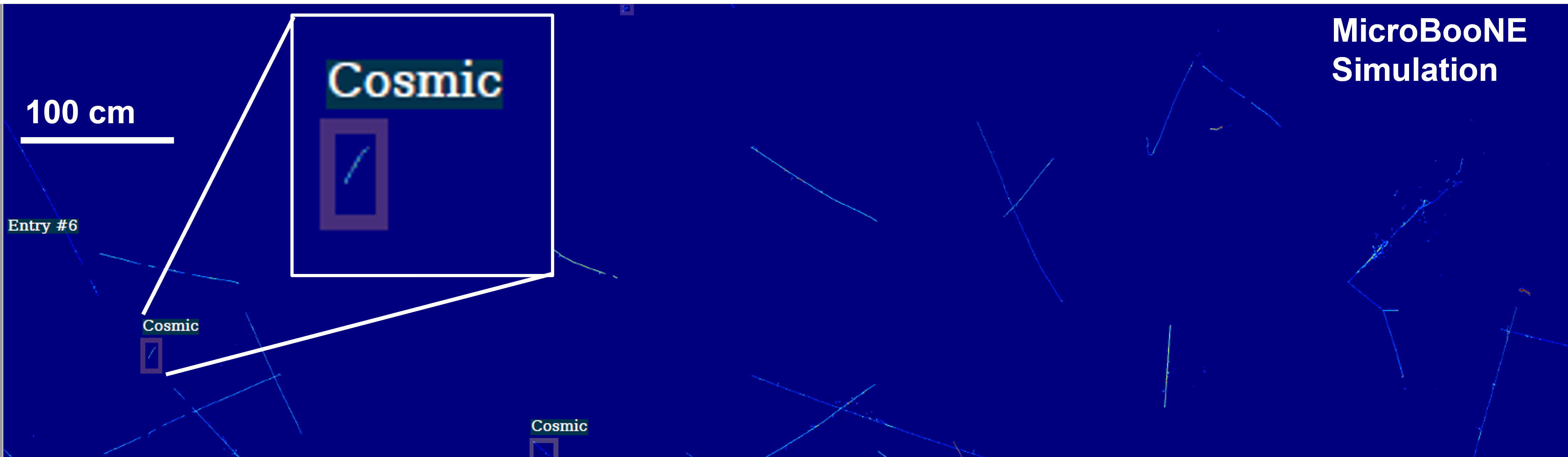}
\caption{Another failure mode for the zero-efficiency peak. Here the cosmic interaction is relatively small in size compared to others in the event image. The colored box is the true interaction, and the white box shows a zoom-in of the area of interest.}
\label{fig:ZeroEff_Garbage}
\end{figure}

The second failure mode that contributes to the zero efficiency peak are true interactions that tend to be smaller in spatial extent compared to a typical simulated interaction, with less charge across fewer pixels in the event image. An example of this is shown in figure \ref{fig:ZeroEff_Garbage}. These interactions are reasonable to expect the network to find as there is nothing to obscure the interaction. However, as they tend to be smaller, they are less likely to overlap with a neutrino interaction in the image or confuse the network, and therefore are a less important part of the background.

\subsection{Interaction coverage}
\label{sec:coverage}

\begin{figure}[h]
\centering
\includegraphics[scale=.2]{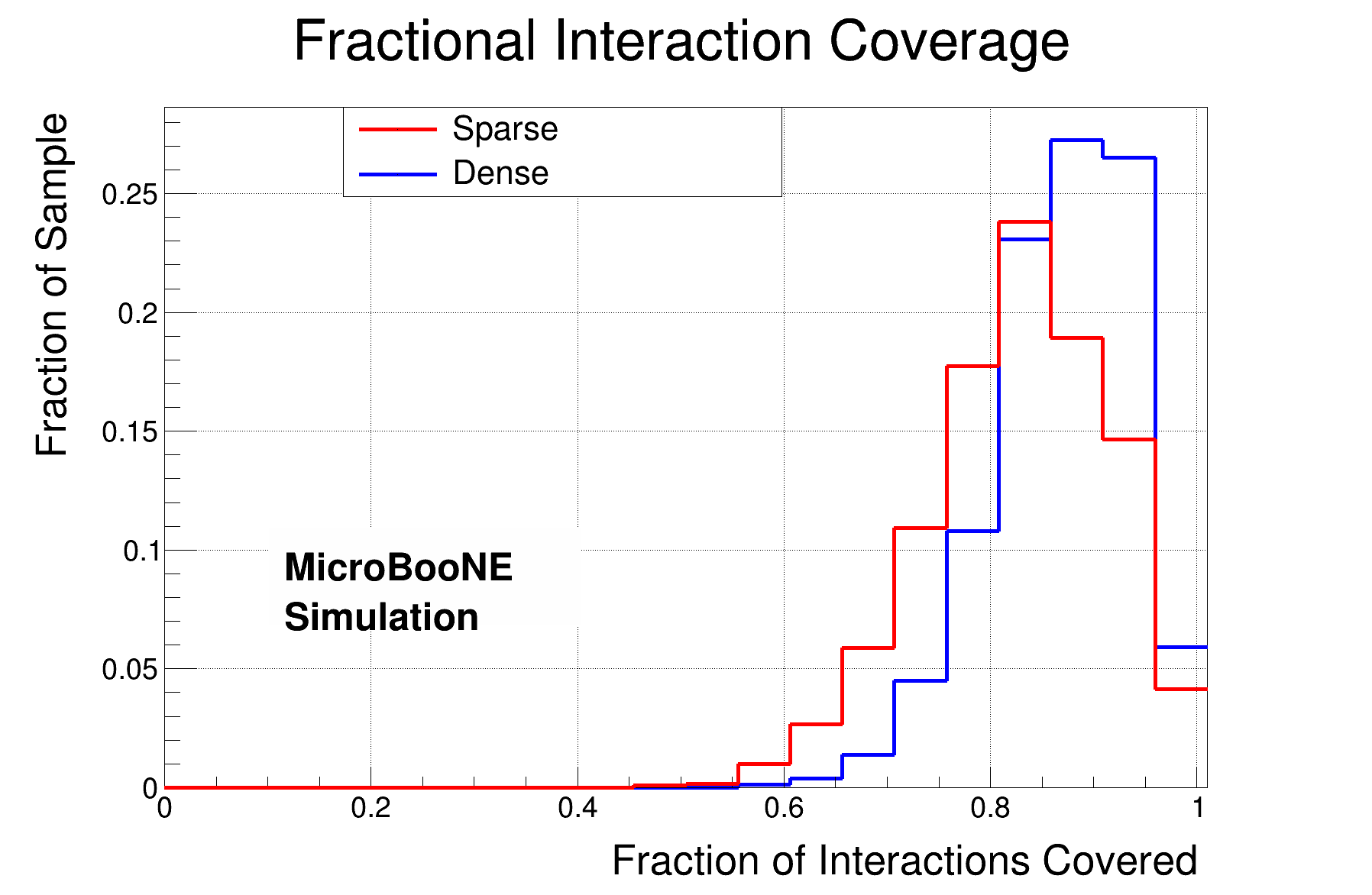}
\caption{The fraction of true interactions in events that have greater than 80\% efficiency as measured on the validation set.}
\label{fig:frac_coverage}
\end{figure}

Now that we have explored a pixel-wise efficiency, we next examine interaction coverage within a given event. We define a true interaction as being "covered" if its pixel-level efficiency as defined in section \ref{sec:effpur} is greater than 80$\%$. This means the network has to cluster the majority of the interaction, while still leaving some room for error. Figure \ref{fig:frac_coverage} compares the fraction of true interactions that are covered in a given event for both the dense and sparse implementations of Mask-RCNN. Note that the fraction of interactions that are "covered" is a metric defined for each event, compared to efficiency and purity, which are pixel-based metrics defined for each true and predicted interaction respectively. Later in section \ref{sec:enufinding} we return to examining the pixel-based efficiency and purity. When examining the fraction of interactions that are covered the dense network has a slight edge over the sparse version, but both networks consistently cover the majority of true interactions within a given event.

\begin{figure}
\centering
\begin{subfigure}{.5\textwidth}
  \centering
  \includegraphics[scale=0.5, width=\linewidth]{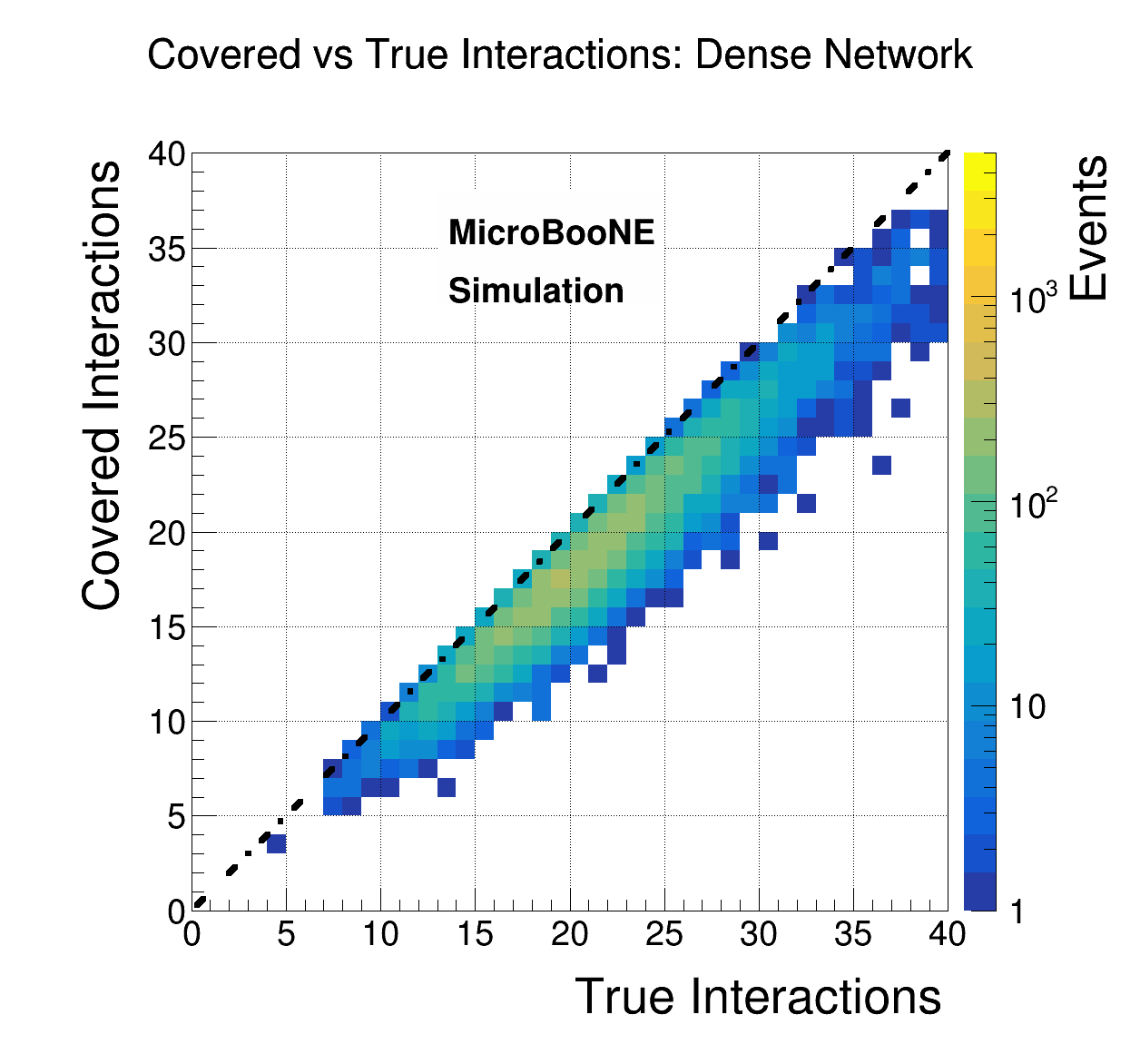}
  \caption{Dense Mask-RCNN}
  \label{fig:Dense_Coverage_2D}
\end{subfigure}%
\begin{subfigure}{.5\textwidth}
  \centering
  \includegraphics[scale=0.5, width=\linewidth]{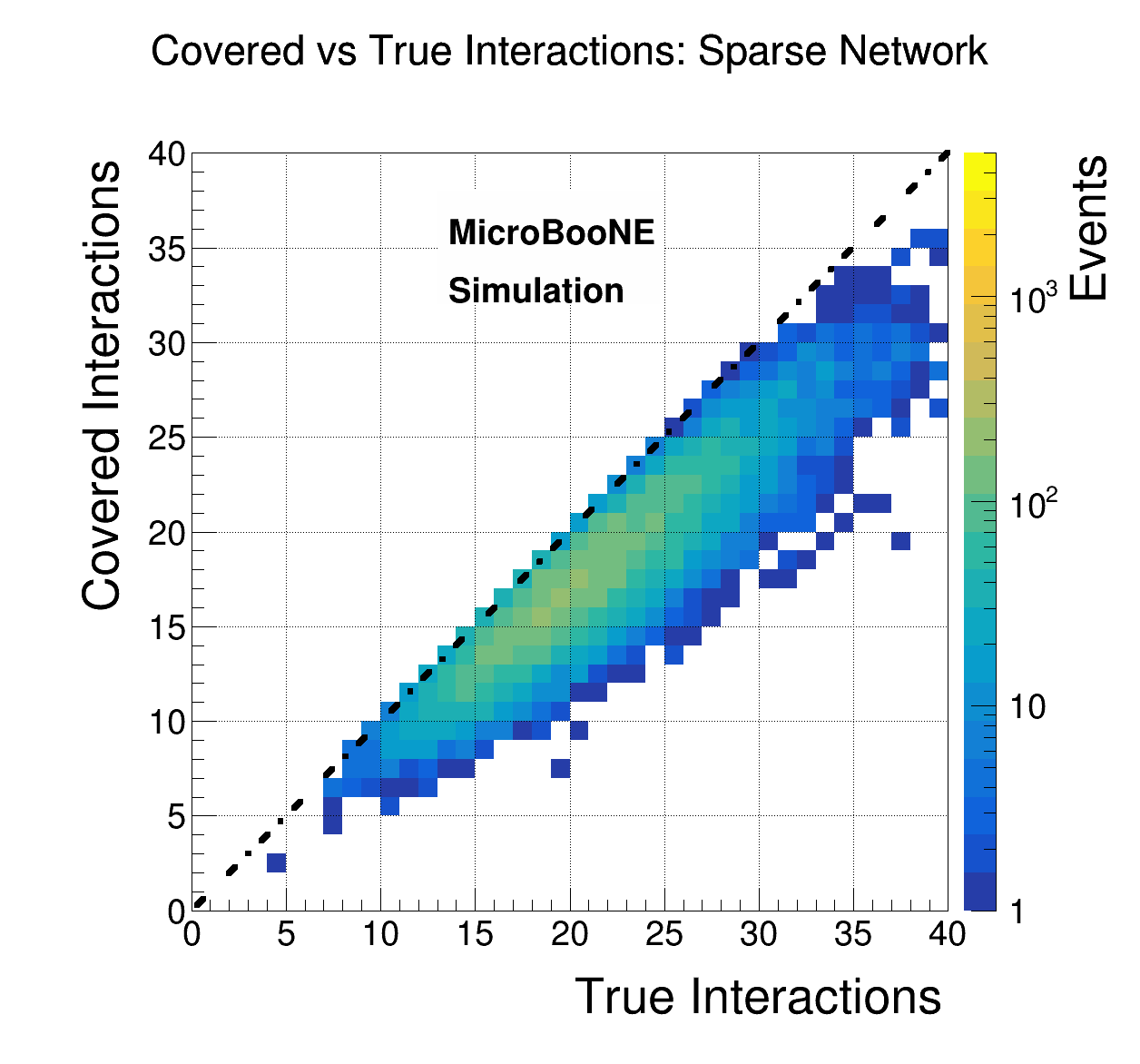}
  \caption{Sparse Mask-RCNN}
  \label{fig:Sparse_Coverage_2D}
\end{subfigure}
\caption{The number of covered interactions is plotted against the number of true interactions. A dashed line along $y=x$ represents the absolute perfect performance, with all true interactions being covered in each event. Out of an average of 20.8 true interactions per event, the dense network covers an average of 18.2, while the sparse network covers 17.2.}
\label{fig:compare_cov2d}
\end{figure}

It is also useful to examine the performance of Mask-RCNN as a function of the number of true interactions in an event. This investigates whether the performance of the network falls off for ‘busier’ events with additional particle interactions cluttering up the image. To examine this, we look at the number of covered true interactions as a function of the number of true interactions in the events. Figure \ref{fig:compare_cov2d} shows this measurement for both the dense and sparse networks.

While the ideal network would cover all of the interactions, we see that both versions of the network produce a distribution slightly below the target line $y=x$. The fact that these distributions are linear demonstrates that the network performance does not diminish as the number of true interactions in a given event increases.

\subsection{Network comparisons discussion}
\label{sec:comparisondiscussion}

In comparing the dense and sparse versions across these various metrics, we find that the dense implementation performs better than the current version of sMask-RCNN. Viewing the two efficiencies of the networks, the peak at 1.0 for the dense Mask-RCNN is larger, and narrower than the peak in the sMask-RCNN distribution.  Similarly the peak at zero-efficiency, representing interactions that are missed, is smaller. The dense Mask-RCNN surpasses sMask-RCNN in average efficiency as well with 89.1$\%$ compared to the slightly lower 85.9$\%$. The dense network covers 87.1$\%$ of interactions compared to the sparse's 82.7$\%$, where "covered" is defined in section \ref{sec:coverage}.

It is important to note that both versions of the network completely miss true interactions with the same frequency. This is shown by observing the peak at 0 efficiency in figure \ref{fig:efficiency_overlaid}. There we see that each network misses about 4$\%$ of interactions. This means that while the metrics point to worse performance for the sparse network, it still finds the interactions themselves, and still covers them to largely the same extent. The difference is that it builds less complete masks of the true interactions compared to the dense network, though it still finds part of them.

We note that it is difficult to track the effects of these differences on the training and learning of the networks. As such, we cannot distinguish whether the difference in performance of the two networks is due to the change from dense ResNet to sparse ResNet, the training on crops versus entire event images, or a combination of both. However, we emphasize that sMask-RCNN's ability to cluster interactions is sufficient for us to compare its cosmic tagging ability to current methods deployed in MicroBooNE, particularly given the speedup acquired by moving to submanifold convolutions. Further, while the dense network's performance does slightly outperform the sparse network, MicroBooNE's data processing prioritizes speed and the use of CPUs in order to scale to the size of its datasets. Therefore, deploying the dense network is not a viable option. This means that regardless of the performance of the dense version of the network, it is prohibitively slow to run at the scale MicroBooNE's dataset requires. Therefore for the analysis performed in section \ref{sec:enufinding} we will only use sMask-RCNN.

\section{Finding electron neutrinos with sMask-RCNN}
\label{sec:enufinding}

In this section, we examine using sMask-RCNN in MicroBooNE to reduce the ratio of cosmic ray background events to electron neutrino events. There are two approaches that we explore, one designed to select neutrino interactions explicitly, and another designed to remove cosmic ray muon interactions. The first, an "identification by positive" approach, would use the neutrino-class output from sMask-RCNN and apply some threshold to select neutrino interactions. This approach is discussed in section \ref{sec:idbypos}. 

The other approach, "identification by negative", applies an event veto, which targets cosmic-only events to flag them for removal. Then the remaining events are those with a neutrino, as well as cosmic ray muons interacting in the detector during the beam window. Finally, we note that while this article specifically targets electron neutrinos, the tools discussed could be trained to target muon neutrinos with the same methodology, then an analogous study could be performed.

Recall that the cosmic ray background in MicroBooNE is very large, and therefore must be dealt with early on in any chain of reconstruction tools. To demonstrate the scale of this problem, we define three different samples:

\begin{enumerate}
  \item \textbf{General Electron Neutrino Sample}: Events containing a simulated electron neutrino interaction combined with cosmic ray muon background data. 
  \item \textbf{Low Energy Electron Neutrino Sample}: The same as sample 1 but only for electron neutrinos with energy less than 400 MeV.
  \item \textbf{Off-Beam Sample}: Data taken by the detector in anti-coincidence with the neutrino beam. This means there is no beam neutrino interaction present. This sample represents the cosmic ray-only background events.
\end{enumerate}

The expected ratios of samples 1 and 2 to sample 3 after current optical and software triggering requirements in the LArTPC are depicted in table \ref{tab:event_freq}. 'Triggering' here represents passing the requirements set by MicroBooNE for data to be written to disk, indicating something of interest occurred during the detector. The rate of electron neutrino events expected in the detector is simulated per protons-on-target (POT). This is a rate of events per protons delivered in spills to the target to create the Booster Neutrino Beam. But by definition, off-beam data will not have any protons-on-target, so a ratio cannot be determined directly. Instead the 'effective' POT for the off-beam sample is determined by taking the ratio of events triggering the LArTPC in the off-beam sample over the number of triggers in an on-beam sample, multiplied by the POT of the on-beam sample.  This 'effective' POT can then be used to determine the ratios in the table. 

These ratios are instructive, as they indicate the initial signal to cosmic ray background event ratio for analyses that seek to remove cosmic ray-only events. These ratios only depict events. Each event contains O(20) cosmic ray muon interactions, and either zero or one neutrino interaction. So the true ratio of cosmic ray muon interactions to electron neutrino interactions is roughly 20 times higher than the ratio of event types. As the purpose of this article is to develop techniques to reduce the significant cosmic ray muon background, we ignore other backgrounds to an electron neutrino signal, such as muon neutrino events.

\begin{table}
\centering
\caption{\label{tab:event_freq}The expected ratio of the two different neutrino sample events to off-beam background events.}
\begin{tabular}{||c  c ||} 
\hline
Type of event & Ratio to off-beam sample events \\ [0.5ex] 
\hline\hline
General electron neutrino &   $5.97\times 10^{-4}$ \\ 
\hline
Low energy electron neutrino &   $1.51\times 10^{-5}$ \\
\hline
\end{tabular}
\end{table}

While the analysis in section \ref{sec:idbypos} uses the validation data used thus far in this article, section \ref{sec:idbyneg} uses the general and low energy electron neutrino samples and cosmic ray-only sample described above. We note that the low energy electron neutrino sample is chosen as it corresponds to the range within which MiniBooNE observes a low-energy-excess \cite{MBExcess}. as The validation data contains only simulated interactions, including CORSIKA-simulated cosmic ray muons. However, the three new datasets contain cosmic ray muons from data, rather than simulation. In previous sections, where we need information about the individual cosmic ray muon interactions for the metrics, it is necessary to use simulated cosmic ray muons. However, this is not the case for the event veto described in section \ref{sec:idbyneg}. Therefore it is better to use cosmic ray interactions from data, as it avoids any potential imperfections in cosmic-ray muon modeling that may be present in simulation. 

\subsection{Electron neutrino identification}
\label{sec:idbypos}

\begin{figure}[h]
\centering
\includegraphics[scale=.2]{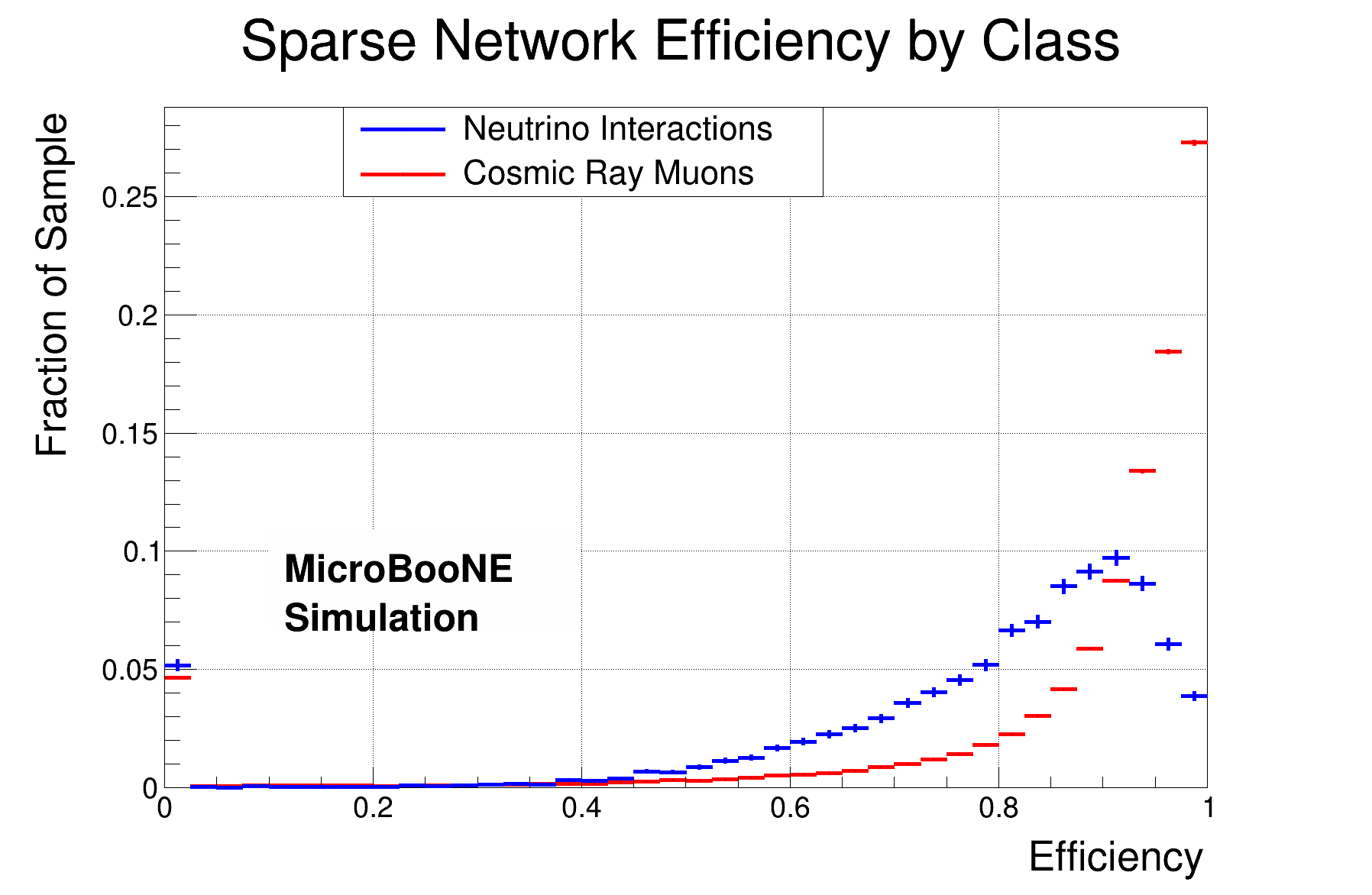}
\caption{The efficiency of sMask-RCNN broken down by class. The average efficiency is 76.8$\%$ for electron neutrinos, and 86.1$\%$ for cosmic ray muons. Statistical uncertainty bars are shown.}
\label{fig:classcompare_efficiency}
\end{figure}

We examine the electron neutrino "identification by positive" approach by looking at the efficiency and purity as defined in section \ref{sec:effpur}, broken down by the two class categories: cosmic ray muons and electron neutrinos. Figure \ref{fig:classcompare_efficiency} shows the efficiency metric (defined in eq.\ \ref{eqn:eff}) for simulated electron neutrino interactions, separated by class. The fraction of neutrino interactions that are 'covered' as defined in section \ref{sec:coverage}, is $59.63\%$. The average efficiency is 76.8$\%$ for electron neutrinos, and 86.1$\%$ for cosmic ray muons. If the network proposes no neutrino interactions, then the efficiency of that event's neutrino interaction is 0. Each class has a peak at 0, but the cosmics also have a peak at 100\%, whereas the neutrino interactions peak at just over 90\%. In the context of neutrino interactions, the interaction has some number of prongs, where a `prong' refers to a shower, or track coming out of the neutrino interaction vertex. It is possible that, for neutrino interactions with at least two prongs, the network fails to mark a shorter prong, or partially masks the track-like portion of an electromagnetic shower with both track- and shower-like topology. Particularly long tracks are also difficult to capture completely due to rescaling within sMask-RCNN, which may lead to the ends of the track getting truncated.

\begin{figure}[h]
\centering
\includegraphics[scale=.2]{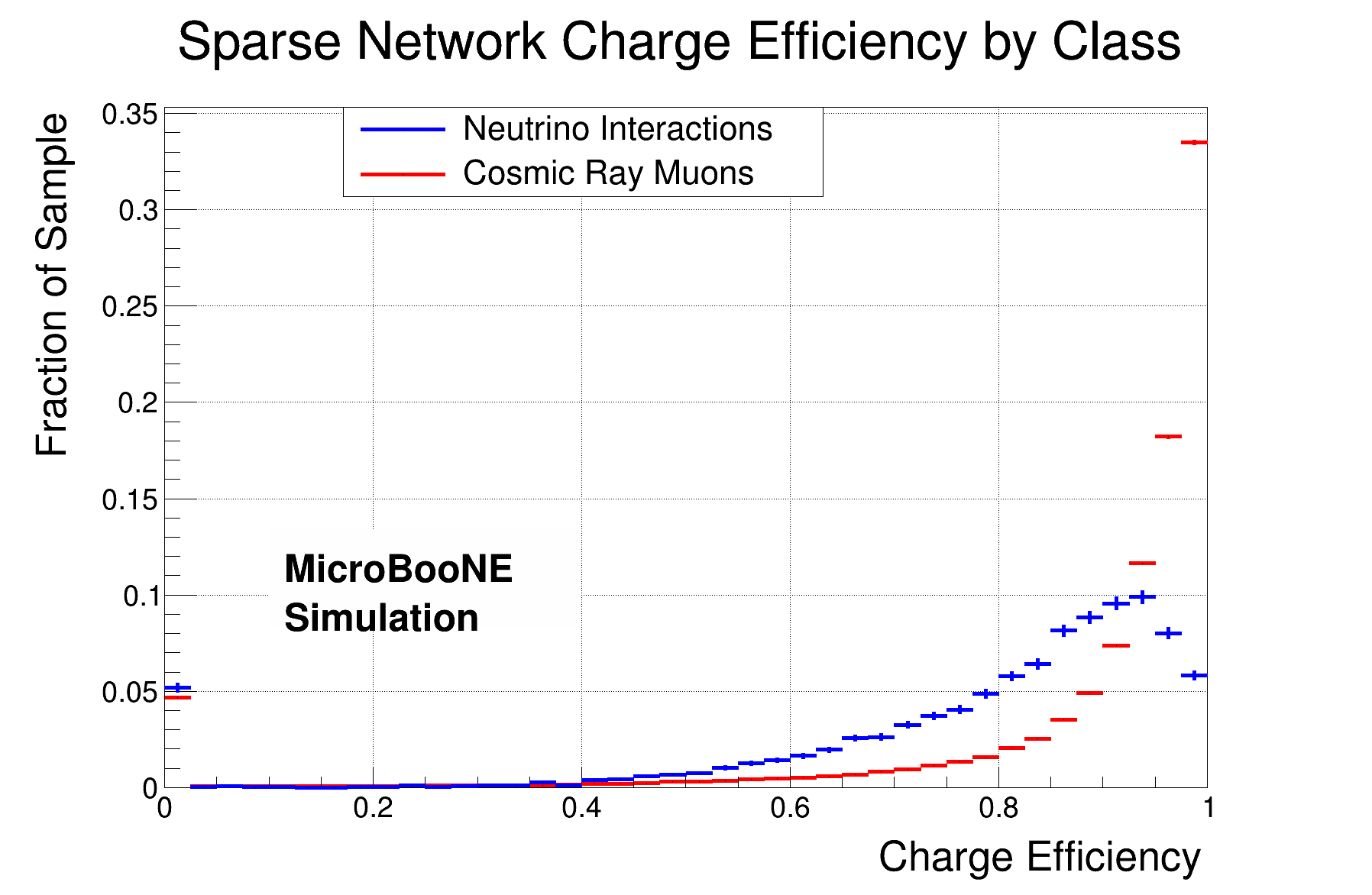}
\caption{The charge efficiency of sMask-RCNN broken down by class. The average charge efficiency for electron neutrinos is 77.9$\%$, and $86.8\%$ for cosmic ray muons. Statistical uncertainty bars are shown.}
\label{fig:classcompare_chargeefficiency}
\end{figure}

Figure \ref{fig:classcompare_chargeefficiency} is also separated by class, but shows the charge efficiency. The average charge efficiency for electron neutrinos is 77.9$\%$, and $86.8\%$ for cosmic ray muons. Both overall and for each class individually the network has a better charge efficiency than standard efficiency, indicating that the interaction masking prioritizes clustering pixels corresponding to larger charge deposition regardless of class. 

We see from these two efficiency breakdowns that the network's ability to find an interaction is not strongly tied to the type of interaction, as the peak at zero efficiency is the same for each class. However the masks for cosmic interactions are more complete than those for neutrino interactions.

\begin{figure}[h]
\centering
\includegraphics[scale=.2]{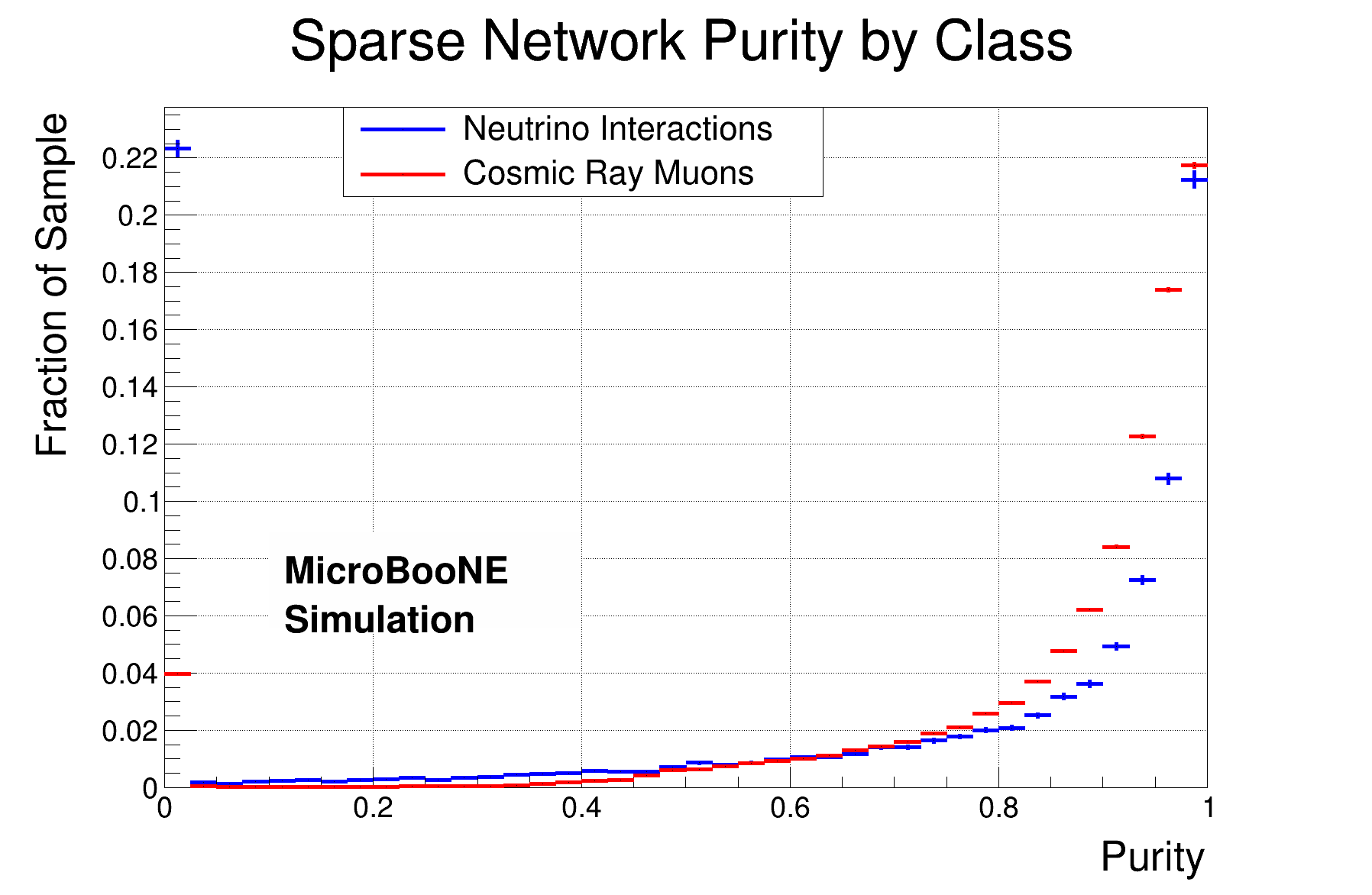}
\caption{The purity of sMask-RCNN broken down by class. Predicted interactions with zero purity are misclassified, for example a true neutrino labeled as a cosmic ray. The average purity is 64.9$\%$ for electron neutrinos and 84.7$\%$ for cosmic ray muons. Statistical uncertainty bars are shown.}
\label{fig:classcompare_purity}
\end{figure}

The purity of sMask-RCNN predicted interactions (defined in eq.\ \ref{eqn:pur}) is broken down by class in figure \ref{fig:classcompare_purity}. The average purity is 64.9$\%$ for electron neutrinos and 84.7$\%$ for cosmic ray muons. Here we see an issue with using this version of sMask-RCNN in an identification by positive approach. The peak at zero purity for the neutrino class indicates that, in events that contain simulated neutrino interactions, roughly 22$\%$ of predicted interactions labeled neutrinos are actually placed on cosmic ray muons. This implies selecting only predicted neutrino interactions yields a ratio of electron neutrinos to cosmic ray muons of 78:22. However, this only applies if the identification were restricted to events that definitely contain a neutrino. When factoring in the significant number of events in the data that contain no neutrino interaction, as indicated by table \ref{tab:event_freq}, the number of falsely identified neutrinos grows much worse.

We can imagine ways to improve this identification by positive approach, from increasing the required confidence score the classifier in sMask-RCNN has in a predicted interaction, to retraining the network with increased penalties for falsely predicting the neutrino class. However, these were set aside in favor of exploring the identification by negative approach in the subsequent section.

 \subsection{Cosmic-only event veto}
\label{sec:idbyneg}
The topology of a muon interaction is much more consistent than that of electron neutrino interactions, due to an electron neutrino interaction's variety of final states. A muon generally creates a track in the detector, although it may potentially create small delta-ray or bremsstrahlung showers along its line of positive charge. The muon may cross out the other side of the detector, stop in the detector (which may produce a Michel electron shower), or – in extremely rare cases – undergo a nuclear interaction in the detector. In comparison, an electron neutrino has many varied final states, with more potential event topologies from different daughter particle scenarios. Most notably, an electron neutrino interaction will lack a long track laid out by a muon, and instead frequently feature a showering particle, such as an electron.

This is consistent with the improved performance of the network labeling cosmic ray muons compared to electron neutrinos, evident in section \ref{sec:idbypos}, where the network must learn to recognize the many different patterns and topologies that make up the electron neutrino class label. Therefore, relying on sMask-RCNN's cosmic ray muon clustering rather than its neutrino clustering may be preferable. As such, the identification by negative approach, which only relies on an understanding of the cosmic ray background, may be more effective.

To study this, we implement an event veto. The goal of this veto is to use sMask-RCNN outputs to separate entire events into those that contain only cosmic ray background, and those that contain an electron neutrino interaction among cosmic rays in the beam window. This task is tested by using this event veto to separate the cosmic ray-only data sample from the general and low energy electron neutrino samples. If this veto were perfect, then the only cosmic ray muons left would be the O(20) interactions per event containing a neutrino interaction. These remaining cosmic ray muons can then be dealt with further down the reconstruction chain. In addition to reducing the background, by flagging entire events an event veto can allow for savings on data storage, and later processing steps. Events labeled as background-only do not need to be saved or further processed while offering the provided signal efficiency.

In order to provide a comparison to current methods used in MicroBooNE, we analyze this event veto using several different versions of cosmic ray tagging. For the first tagger, we include all cosmic ray interaction pixels predicted by sMask-RCNN with a confidence score greater than 0.20. Reducing the confidence score requirement relative to earlier sections of this article allows more cosmic removal at the expense of including multiple overlapping cosmic interaction predictions.  Decreasing the confidence score requirement cannot negatively impact the electron neutrino efficiency, as it only serves to include strictly more predictions from the network. For individual interaction labeling, shown earlier in the article, this would be problematic. However, for the event veto described below, we are concerned with removing entire events.

For the next tagger, we add the pixels tagged as cosmic ray muons by MicroBooNE's Wire-Cell Q-L described in section \ref{sec:existingtools}. This adds information from the PMT light collection system and the two LArTPC induction planes, none of which is used by sMask-RCNN. Therefore by comparing sMask-RCNN alone to this combined tagger, we can see the additional value provided by the light information, as well as demonstrate the effectiveness of sMask-RCNN operating in a regime with less information. Recall WC Q-L matching is a piece of the full WC cosmic tagger. We repeat the following analysis with the full WC cosmic tagger at the end of this section, exploring two additional tagger configurations.

A perfect tagger would tag every pixel containing deposited charge associated with cosmic ray muons. All that would remain in the event image would be pixels holding charge corresponding to a neutrino interaction, if present. However, with the expectation of imperfect performance, we re-cluster the untagged pixels via a "density-based spatial clustering of applications with noise" (DBScan) algorithm \cite{DBScanBOOK}. This means that first we perform our cosmic ray muon tagging, remove those pixels from the image, then run DBScan on the resulting image. DBScan is chosen for this reconstruction task to provide a handle on the size of remaining clusters in the image after the cosmics have been removed. When used on data this will provide us a handle as to how large of clusters remain behind after initial cosmic tagging.

\begin{figure}
\centering
\begin{subfigure}{.5\textwidth}
  \centering
  \includegraphics[scale=0.5, width=\linewidth]{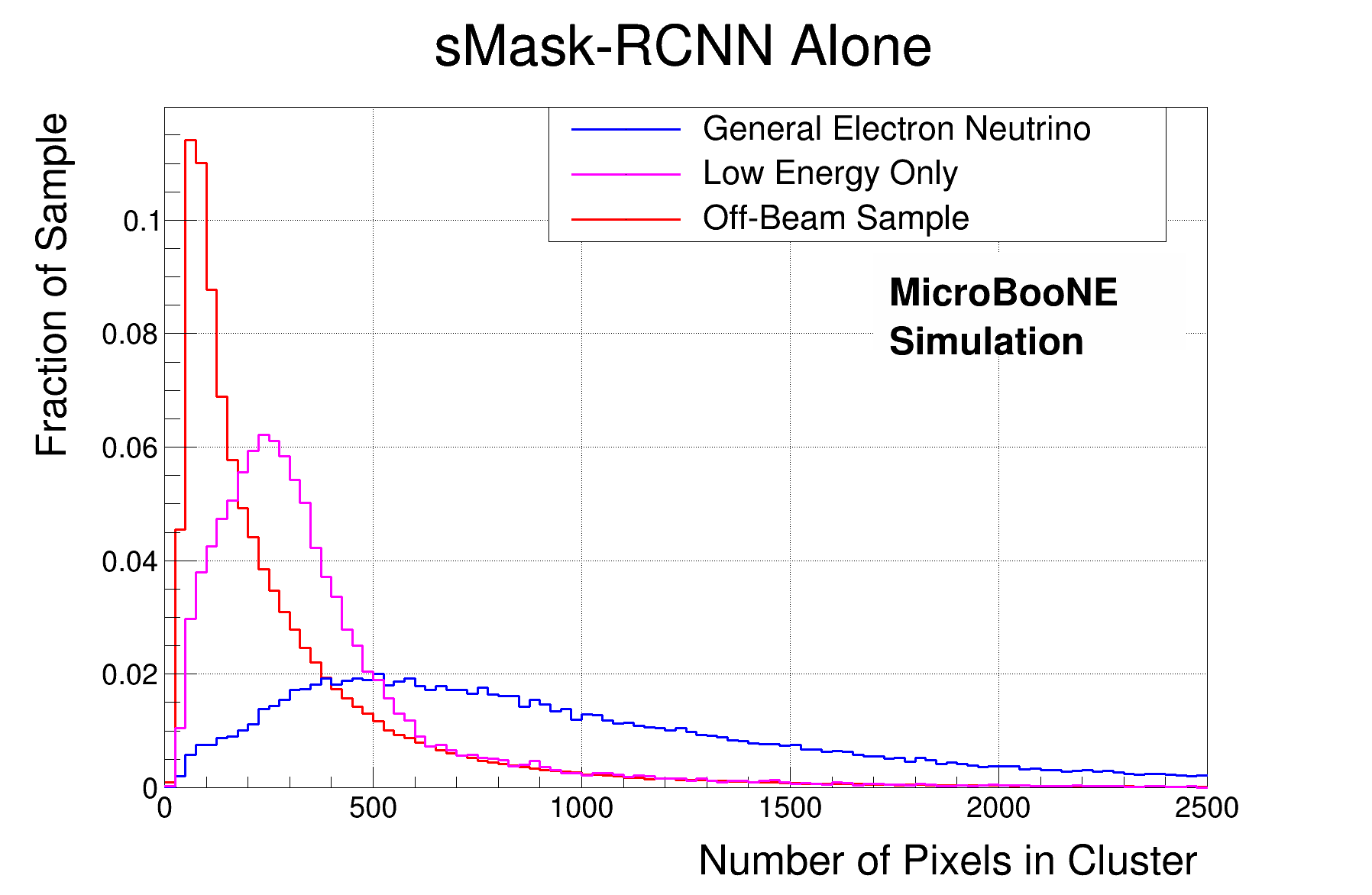}
  \caption{Cosmic tagging by sMask-RCNN alone}
  \label{fig:bigclustsMask}
\end{subfigure}%
\begin{subfigure}{.5\textwidth}
  \centering
  \includegraphics[scale=0.5, width=\linewidth]{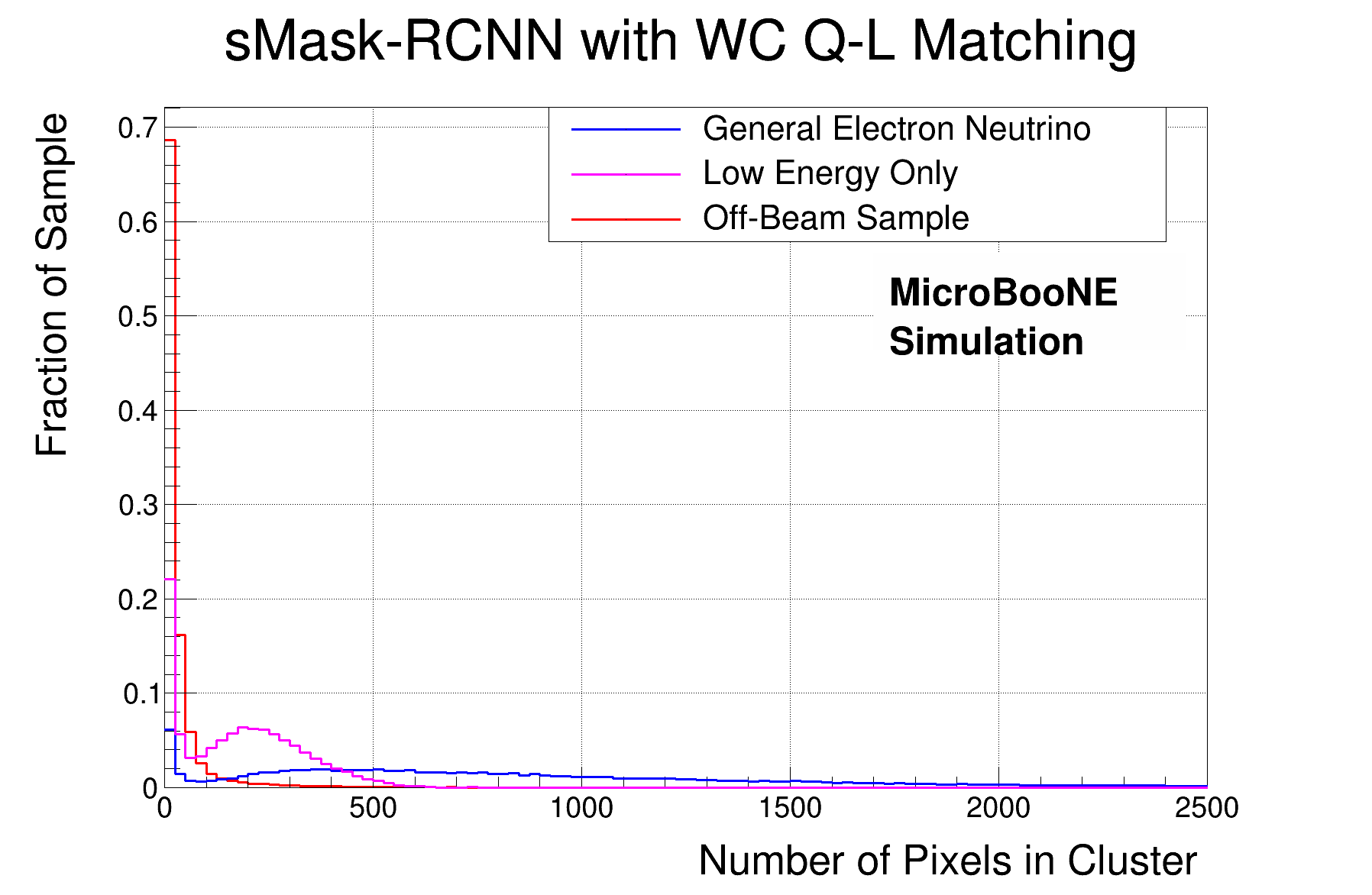}
  \caption{With WC Q-L matching}
  \label{fig:bigclustsMaskWCCL}
\end{subfigure}
\caption{The size of the largest cluster found by DBScan after cosmic tagging.}
\label{fig:bigclustmaskwccl}
\end{figure}

DBScan will output clusters of remaining pixels for each event. These pertain to portions of cosmic ray muon interactions not fully tagged, and the neutrino interaction if present.  In the case where both neutrino and cosmic ray muon clusters are present, one large cluster will usually represent the neutrino interaction and several smaller clusters represent the untagged parts of muons. This means the size of the largest cluster is a metric that we can use to isolate events containing neutrino interactions. Figure \ref{fig:bigclustmaskwccl} shows the size of the largest of these clusters for sMask-RCNN with and without the WC Q-L matching algorithm. Each figure shows the distribution of the three key samples described above.

Examining the distributions of these three samples, we see a difference between the off-beam sample, which represents the cosmic ray-only event background, and the two electron neutrino samples. Notably, the off-beam sample generally has a smaller number of pixels in the largest cluster found by DBScan. In figure \ref{fig:bigclustsMask}, we see that the sMask-RCNN tagger produces a distinctly shaped distribution for each sample. The off-beam sample has a peak closest to zero, the low energy sample is shifted slightly to the right, and the general electron neutrino sample is shifted further. This reinforces the expectation, as the neutrino pixels remain in the image untagged, allowing for DBScan to find larger clusters. Examining sMask-RCNN with the WC Q-L matching algorithm in figure \ref{fig:bigclustsMaskWCCL}, we see a shift to the left in all three distributions, placing a strong peak at zero, indicating that the combined tagger frequently labels more pixels as belonging to cosmic ray muon interactions than the sMask-RCNN-based tagger alone. This is particularly notable in the case of the off-beam sample, where the peak at zero accounts for almost 70$\%$ of the sample, compared to about 16$\%$ of the low energy sample, and about 6$\%$ of the general electron neutrino sample. 

Using the size of the largest remaining cluster, we can create a receiver operating characteristic (ROC) curve to demonstrate the efficiency and rejection power of the different cosmic tagging methods when applied as an event veto. Ideally off-beam events have a largest cluster size of 0, with all pixels being removed by the cosmic tagger, while events containing a neutrino retain a large cluster of pixels associated with the neutrino interaction. A ROC curve is a measure of signal retention or signal efficiency on one axis, and background rejection on the other. A curve is created by incrementing some requirement on events. Our requirement is on the size of the largest cluster found by DBScan, described below. Increasing this requirement slowly decreases signal retention and increases  background rejection. Ideally the curve has points in the upper right region of the plot such that signal retention and background rejection are both high.

\begin{figure}
\centering
\begin{subfigure}{.5\textwidth}
  \centering
  \includegraphics[scale=0.5, width=\linewidth]{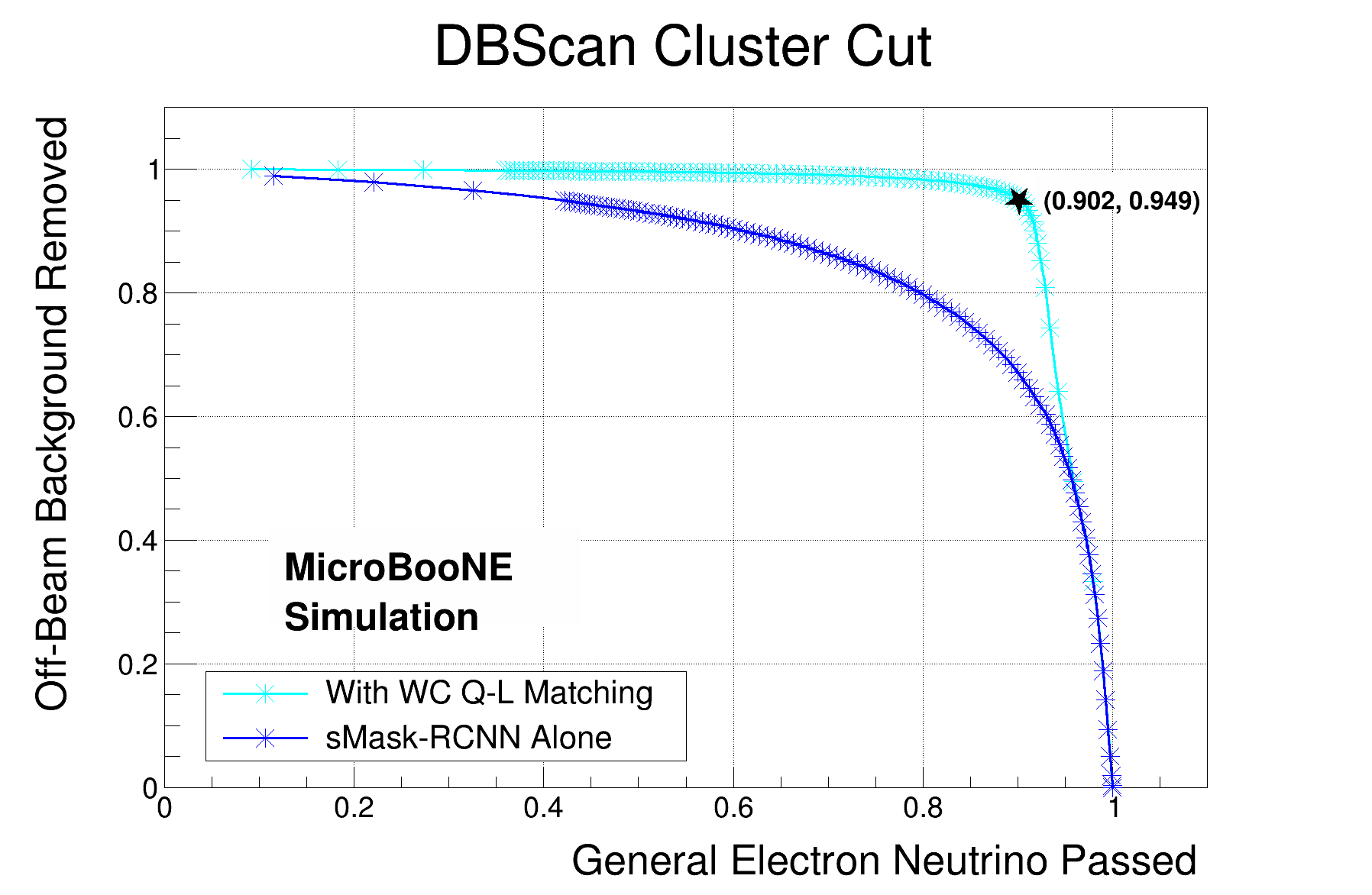}
  \caption{General electron neutrino sample}
  \label{fig:rocgenWCCL}
\end{subfigure}%
\begin{subfigure}{.5\textwidth}
  \centering
  \includegraphics[scale=0.5, width=\linewidth]{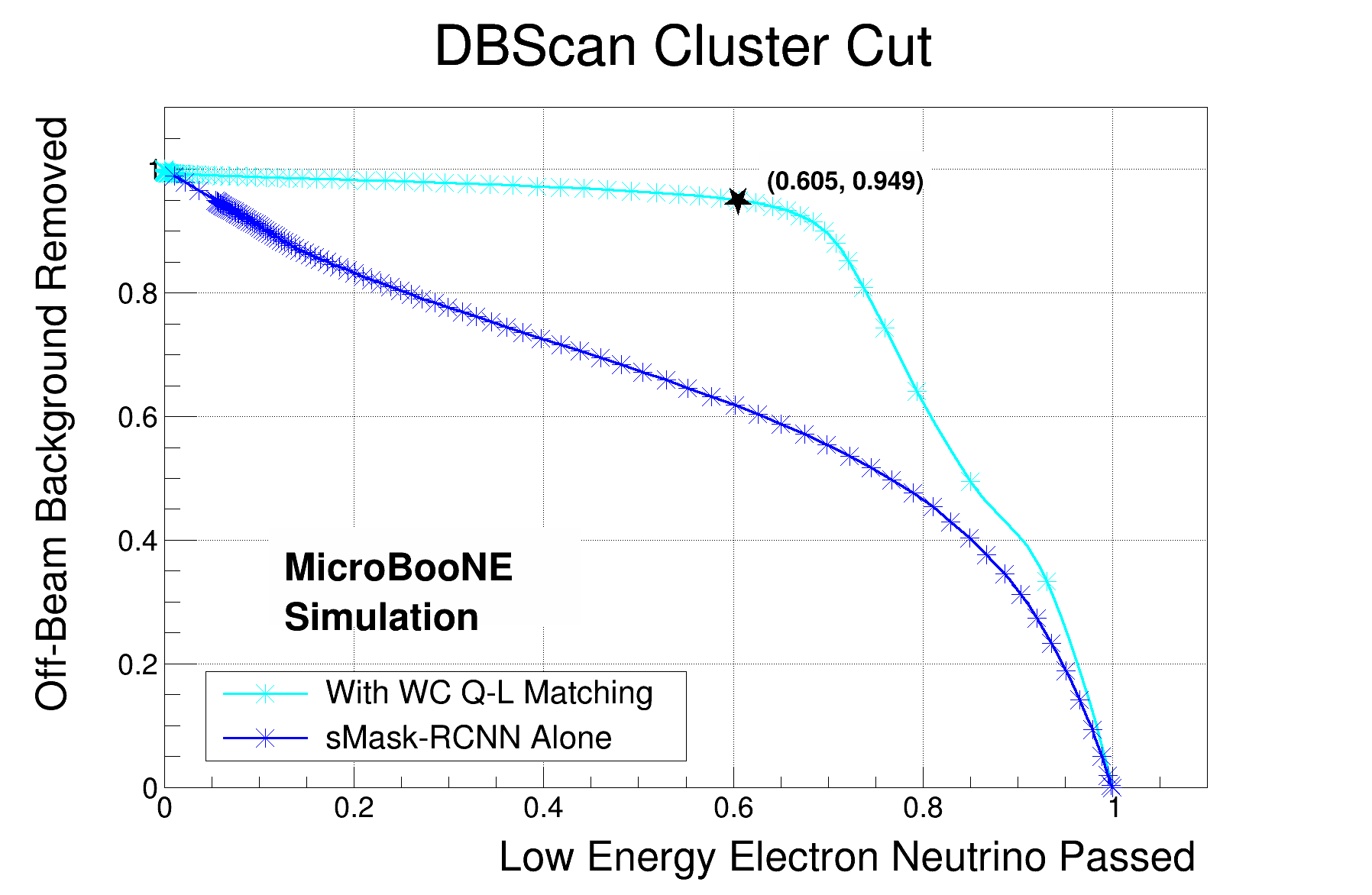}
  \caption{Low energy electron neutrino sample}
  \label{fig:roclowWCCL}
\end{subfigure}
\caption{ROC curves for the sMask-RCNN with and without WC Q-L matching based on a requirement on largest cluster size. Curves are shown for the two electron neutrino samples against the off-beam background.}
\label{fig:rocWCCL}
\end{figure}

We reiterate that we want to explore this quantity  --- largest remaining cluster --- as an event-level discriminant. As such, it will be the value we increment to create ROC curves. Specifically, these curves are made by applying a requirement to filter events that do not contain a DBScan cluster of size greater than $X$, where $X$ is incremented from no requirement, to 0, and then incremented by 10 pixels thereafter. The resulting signal retention and background rejection rates give points for the curves. In figure \ref{fig:rocWCCL}, we show the ROC curves for the signal retentions of the general and low energy electron neutrino samples versus the rejection of the off-beam cosmic ray muon background sample. Curves are made for the two different tagging methods.  We observe that the combined version of a cosmic tagger using both sMask-RCNN and WC Q-L matching yields a better combination of signal efficiency and background rejection. 

For the low energy sample each tagging method performs worse compared to the general sample. However, this is not surprising as the lower energy electron neutrino interactions correspond to less charge deposited in the event image and fewer neutrino pixels in the event. This means that the remaining DBScan clusters related to the lower energy neutrino interactions will be smaller, and harder to isolate from the untagged off-beam cosmic ray muon sample's distribution.

Examining the ROC curve for sMask-RCNN with WC Q-L matching in figure \ref{fig:rocgenWCCL}, we are able to achieve a general electron neutrino signal efficiency of 90.2\% while rejecting 94.9\% of the off-beam cosmic background if we remove events that do not have a cluster of at least 130 pixels after the taggers are run. For the low energy electron neutrino sample in figure \ref{fig:roclowWCCL} we can achieve a signal efficiency of 60.5\% for the same requirement, though a reduced requirement on remaining cluster size could be applied to increase the efficiency at the cost of rejection power, as indicated by the combined curve.

\begin{figure}
\centering
\begin{subfigure}{.5\textwidth}
  \centering
  \includegraphics[scale=0.5, width=\linewidth]{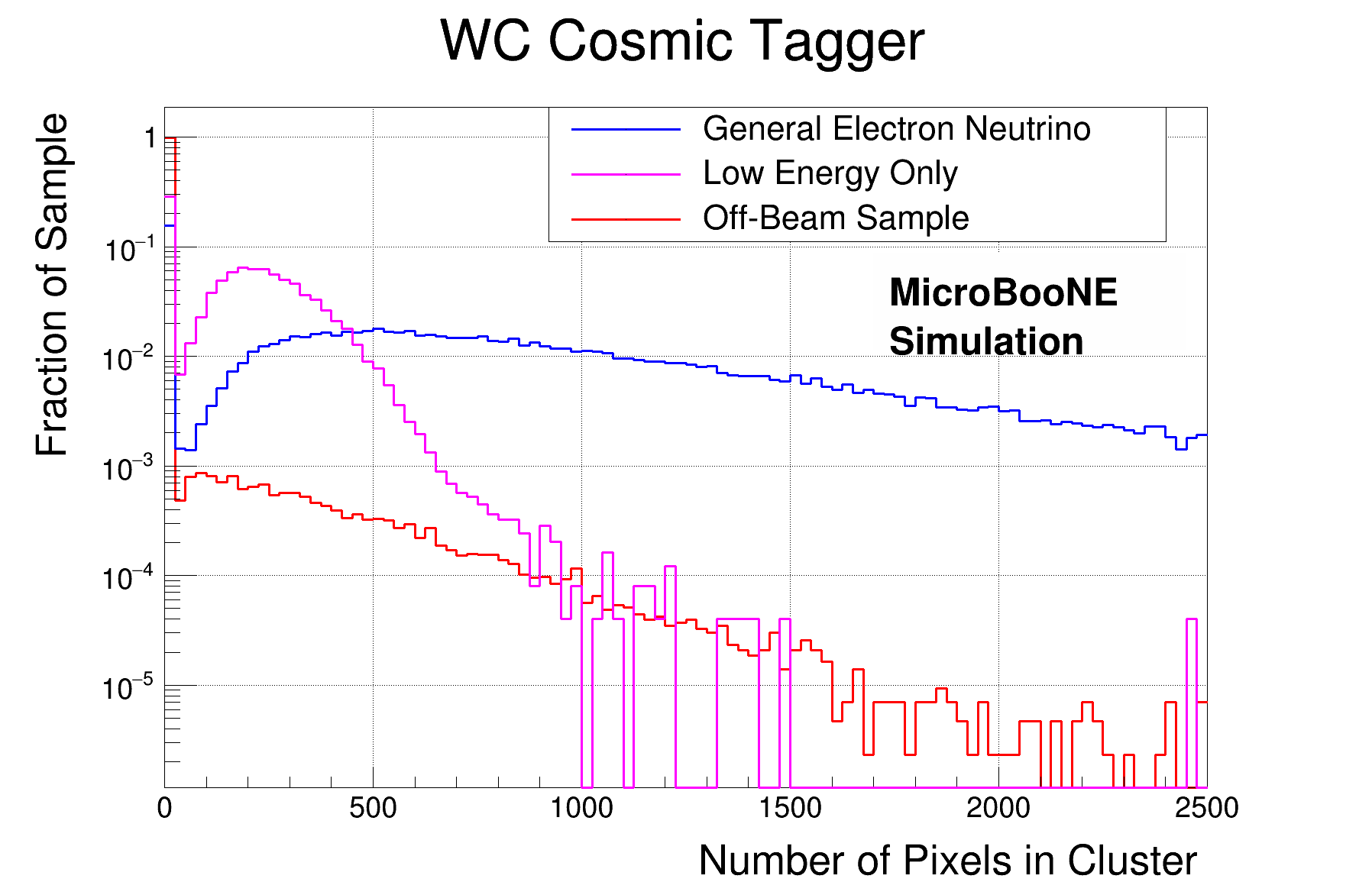}
  \caption{Cosmic tagging by the WC cosmic tagger alone}
  \label{fig:bigclustwcfull}
\end{subfigure}%
\begin{subfigure}{.5\textwidth}
  \centering
  \includegraphics[scale=0.5, width=\linewidth]{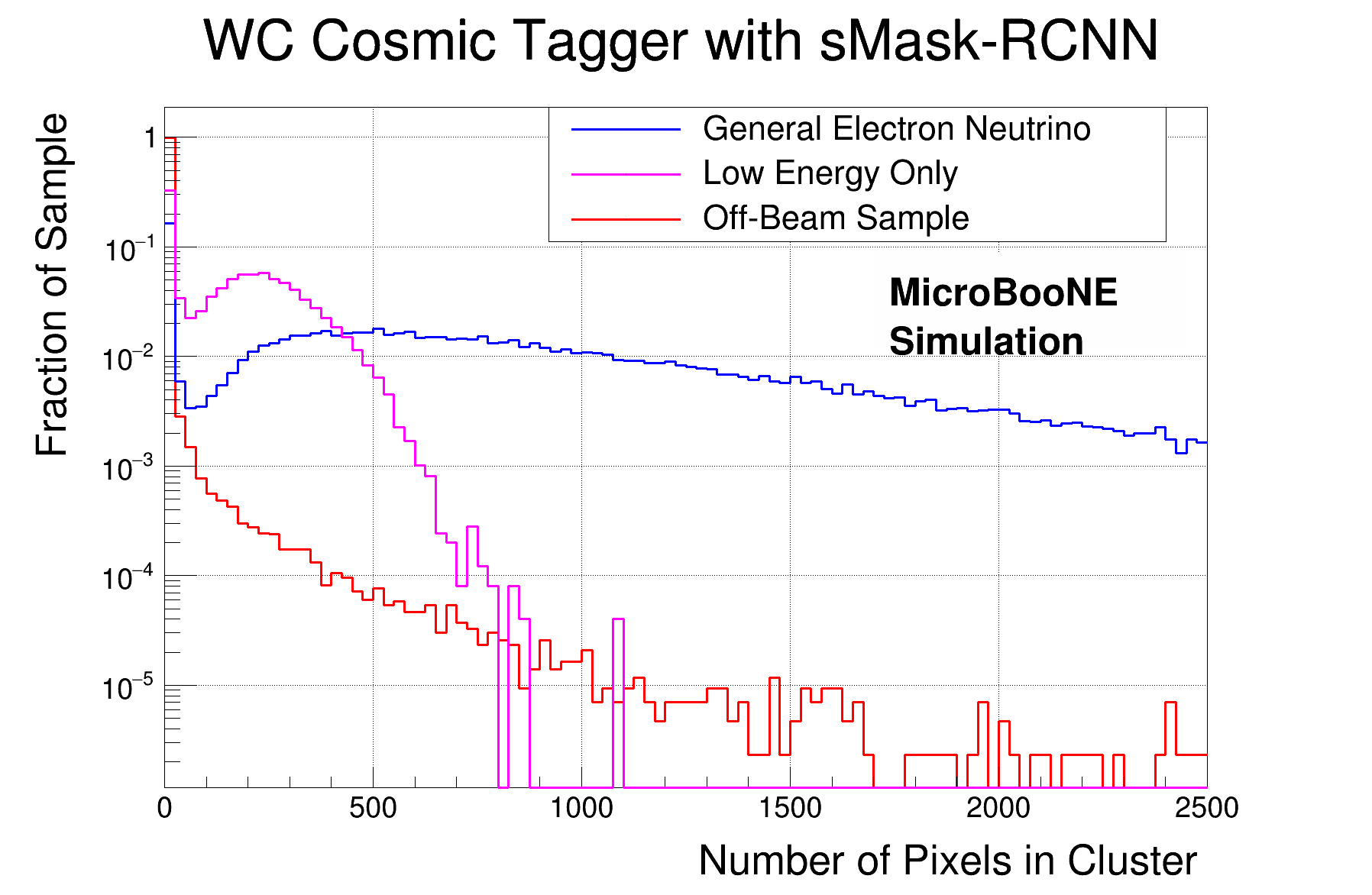}
  \caption{With sMask-RCNN}
  \label{fig:bigclustwcfullMask}
\end{subfigure}
\caption{The size of the largest cluster found by DBScan after cosmic tagging. Note the logarithmic scale.}
\label{fig:bigclustmaskwcfull}
\end{figure}

We also examine the effect of adding sMask-RCNN's cosmic finding to the state-of-the-art complete WC cosmic tagger. This means that we take the event vetoes and cosmic tagging of the WC cosmic tagger, and add the cosmic tagging of sMask-RCNN to get a combined tagger. In figure \ref{fig:bigclustmaskwcfull} we show the distributions of the largest cluster found by DBScan after running the WC cosmic tagger with and without the cosmic ray muons found by sMask-RCNN. For the events that are rejected by one of the WC cosmic tagger event vetos, the largest DBScan cluster is defined to be zero.

The distributions of cluster size before and after adding sMask-RCNN to the WC cosmic tagger show a shift to the left in the shape of the off-beam sample, indicating the added value of sMask-RCNN in cosmic tagging. Without sMask-RCNN, there appears to be a slight peak beyond zero that gets shifted to zero after sMask-RCNN is added. We observe minimal shift in the two electron neutrino samples and each distribution still has a clear second peak separate from zero.

\begin{figure}
\centering
\begin{subfigure}{.5\textwidth}
  \centering
  \includegraphics[scale=0.5, width=\linewidth]{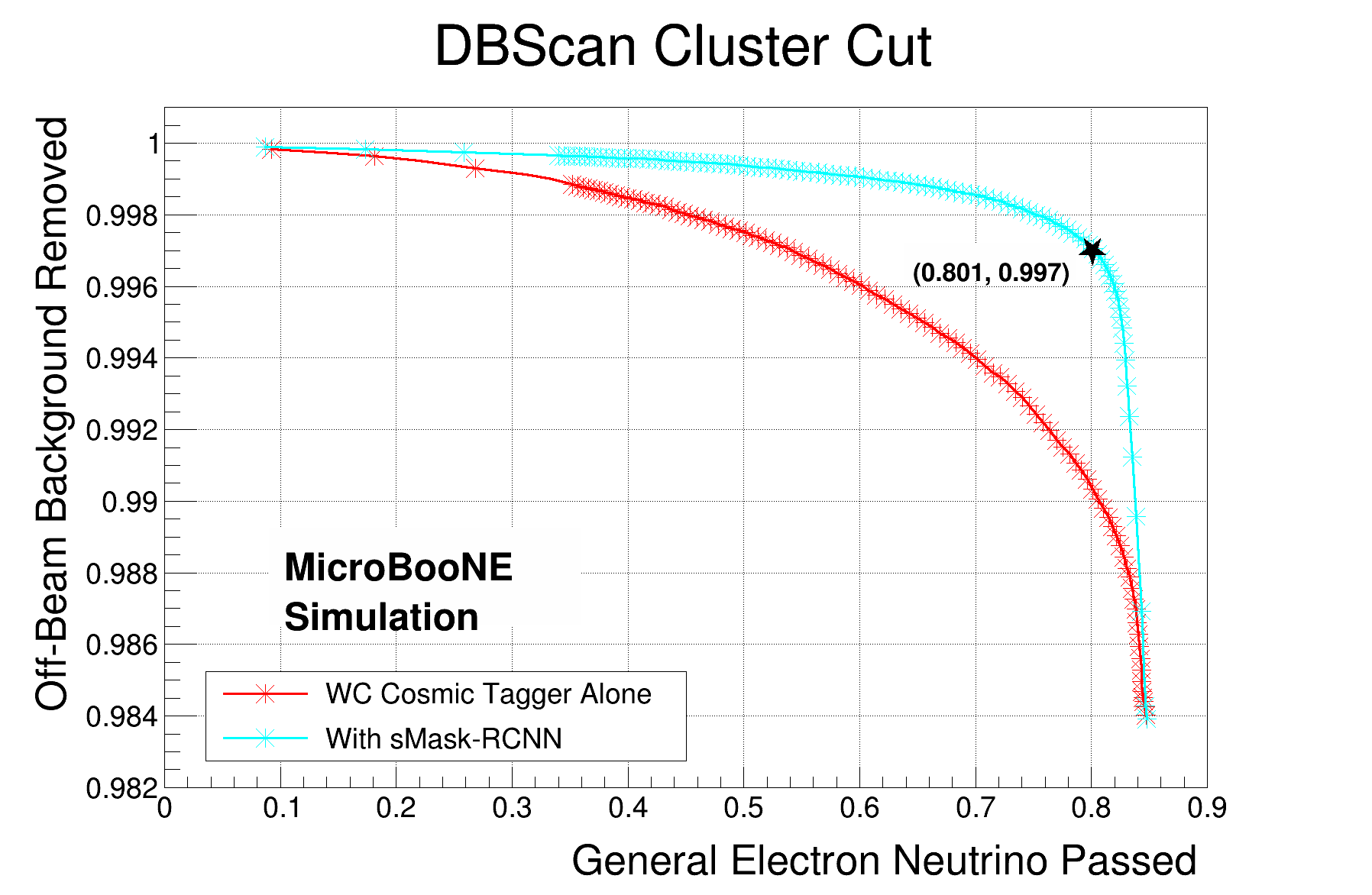}
  \caption{General electron neutrino sample}
  \label{fig:rocgenWCFull}
\end{subfigure}%
\begin{subfigure}{.5\textwidth}
  \centering
  \includegraphics[scale=0.5, width=\linewidth]{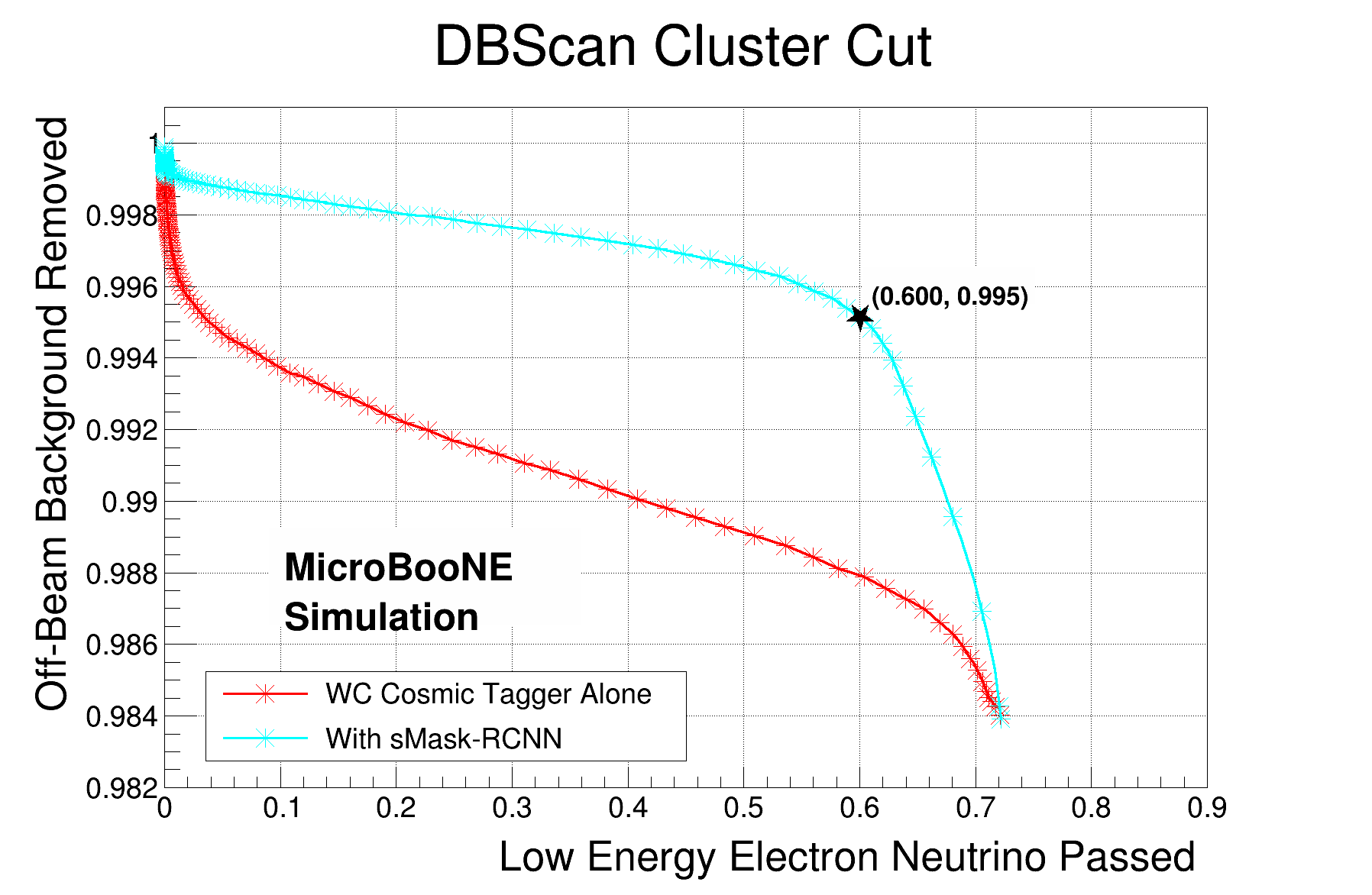}
  \caption{Low energy electron neutrino sample}
  \label{fig:roclowWCFull}
\end{subfigure}
\caption{ROC curves for the WC cosmic tagger with and without sMask-RCNN based on a requirement on largest cluster size. Curves are shown for the two electron neutrino samples against the off-beam background. Note the suppressed $y$-axis shown here demonstrates significant improvement in background removal compared to figure \ref{fig:rocWCCL}.}
\label{fig:rocWCFull}
\end{figure}

ROC curves for the WC cosmic tagger with and without sMask-RCNN are shown in figure \ref{fig:rocWCFull}. However, as the WC cosmic taggers introduces several of its own event vetos, the point referring to the loosest cut, with the most signal passed, does not allow all events through the veto. Instead it starts with the signal efficiency and background rejection of the WC cosmic tagger, and adjusts as we increase the strength of the DBScan cluster size requirement. We reiterate that the difference between figures \ref{fig:rocWCFull} and \ref{fig:rocWCCL} is the additional event vetos added to WC Q-L matching to create the WC cosmic tagger.

Examining these ROC curves we see that including sMask-RCNN on top of the WC cosmic tagger does yield improvement to the background rejection at equivalent signal efficiencies. In the general electron neutrino sample, for a signal efficiency of 80.1$\%$ the WC cosmic tagger rejects 99.0$\%$ of the background, whereas adding sMask-RCNN rejects 99.7$\%$ of the background at the same signal efficiency. This represents a reduction of the remaining background by 70$\%$. For the low energy neutrino sample, a similar effect is seen, albeit at lower signal efficiencies.

In order to evaluate the difference between these two rejection strengths, it is important to recall the imbalance between off-beam background events and electron neutrino signal events described in table \ref{tab:event_freq}. The 99.0$\%$ rejection provided by the WC cosmic tagger corresponds to a general electron neutrino signal to off-beam background ratio of 1.26, a vast improvement over the starting ratio in the table. However, the addition of sMask-RCNN increases this signal to background ratio to 4.14 by improving the rejection power to 99.7$\%$. For this same selection, the low energy signal to background ratio is 0.015 with the WC cosmic tagger, and 0.56 with after the addition of sMask-RCNN.

\section{Conclusions}
\label{sec:conclusions}
This article demonstrates a novel approach to cosmic ray muon tagging using sMask-RCNN. We demonstrate the ability of this network to locate, identify, and cluster particle interactions in the MicroBooNE LArTPC. We analyze the ability to cluster both the topologically simple cosmic ray muon interactions, as well as highly varied electron neutrino interactions. We also isolate potential failures of the network as areas to improve, largely occurring in interactions appearing heavily unresponsive regions of the LArTPC, or visually small interactions within a 2D wire-plane image.

We modify the original Mask-RCNN framework by substituting sparse submanifold convolutions in the ResNet portion of the network to create sMask-RCNN. Due to the low pixel occupancy of MicroBooNE event image data this leads to a $20\times$ speedup in ResNet processing time on a CPU, as well as decreased runtime memory usage. This improvement is critical in allowing sMask-RCNN to be deployed as a reconstruction tool on CPU farms to scale to high volume data samples that particle physics experiments typically employ. 

This analysis also includes several versions of an event veto. The strongest of these demonstrates that adding sMask-RCNN to the state-of-the-art WC cosmic tagger which is currently used in MicroBooNE reduces the cosmic ray-only event background by a further 70$\%$ and increases the signal to background ratio of electron neutrino events to cosmic ray-only events by more than a factor of three. This means that application of this technique to future measurements in MicroBooNE will result in improvements over current MicroBooNE reconstruction.

\acknowledgments
This document was prepared by the MicroBooNE collaboration using the
resources of the Fermi National Accelerator Laboratory (Fermilab), a
U.S. Department of Energy, Office of Science, HEP User Facility.
Fermilab is managed by Fermi Research Alliance, LLC (FRA), acting
under Contract No. DE-AC02-07CH11359.  MicroBooNE is supported by the
following: the U.S. Department of Energy, Office of Science, Offices
of High Energy Physics and Nuclear Physics; the U.S. National Science
Foundation; the Swiss National Science Foundation; the Science and
Technology Facilities Council (STFC), part of the United Kingdom Research 
and Innovation; the Royal Society (United Kingdom); and The European 
Union’s Horizon 2020 Marie Sklodowska-Curie Actions. Additional support 
for the laser calibration system and cosmic ray tagger was provided by 
the Albert Einstein Center for Fundamental Physics, Bern, Switzerland. 
We also acknowledge the contributions of technical and scientific staff 
to the design, construction, and operation of the MicroBooNE detector 
as well as the contributions of past collaborators to the development 
of MicroBooNE analyses, without whom this work would not have been 
possible.

\section{Bibliography}

%

\end{document}